\documentclass[pre,floats,aps,superscriptaddress,showpacs,nofootinbib]{revtex4}
\usepackage{epsfig}
\usepackage{graphicx}
\usepackage{amssymb, amsmath}
\usepackage{float}
\newcommand{\ddr}{\textrm{d}}
\newcommand{\dd}{\textrm{d}}
\newcommand{\ee}{\text{e}}
\newcommand{\bv}{{\bf v}}
\newcommand{\bc}{{\bf c}}
\newcommand{\ba}{{\bf a}}
\newcommand{\bn}{{\bf n}}
\newcommand{\bF}{{\bf F}}

\newcommand{\bk}{{\bf k}}
\newcommand{\mut}{{\tilde \mu}}
\newcommand{\lambt}{{\tilde \lambda}}
\newcommand{\lambdat}{{\tilde \lambda}}
\newcommand{\vt}{{\tilde v}_0}
\newcommand{\ft}{{\tilde f}}
\newcommand{\pit}{{\tilde \pi}}
\newcommand{\wt}{{\tilde w}}
\newcommand{\cWt}{{\tilde {\cal W}}}
\newcommand{\cW}{{\cal W}}
\newcommand{\cD}{{\cal D}}  
\newcommand{\cP}{{\cal P}}

\newcommand{\cO}{{\cal O}}
\newcommand{\cS}{{\cal S}}
\newcommand{\bsig}{\hat{\boldsymbol \sigma}}

\newcommand{\Mvariable}{{}}
\newcommand{\p}{\partial}
\graphicspath{{figures/}}

\begin{document}
\title{Fluctuations of power injection in randomly driven granular gases.}
\date{\today}
\author{Paolo Visco}
\affiliation{Laboratoire de Physique Th\'eorique (CNRS UMR 8627), B\^atiment
  210, Universit\'e Paris-Sud, 91405 Orsay cedex, France}
\affiliation{Laboratoire de Physique Th\'eorique
et Mod\`eles Statistiques (CNRS
UMR 8626), B\^atiment 100, Universit\'e Paris-Sud, 91405 Orsay cedex, France}

\author{Andrea Puglisi}
\affiliation{Laboratoire de Physique Th\'eorique (CNRS UMR8627), B\^atiment
  210, Universit\'e Paris-Sud, 91405 Orsay cedex, France}

\author{Alain Barrat}
\affiliation{Laboratoire de Physique Th\'eorique (CNRS UMR8627), B\^atiment
  210, Universit\'e Paris-Sud, 91405 Orsay cedex, France}

\author{Emmanuel Trizac}
\affiliation{Laboratoire de Physique Th\'eorique
et Mod\`eles Statistiques (CNRS
UMR 8626), B\^atiment 100, Universit\'e Paris-Sud, 91405 Orsay cedex, France}

\author{Fr\'ed\'eric van Wijland}
\affiliation{Laboratoire Mati\`ere et Syst\`emes Complexes 
(CNRS UMR 7057), Universit\'e
Denis Diderot (Paris VII), 2 place Jussieu, 75251 Paris cedex 05, France}
\affiliation{Laboratoire de Physique Th\'eorique (CNRS UMR8627), B\^atiment
  210, Universit\'e Paris-Sud, 91405 Orsay cedex, France}

\begin{abstract}
  We investigate the large deviation function $\pi_{\infty}(w)$ for the
  fluctuations of the power ${\cal W}(t)=w t$, integrated over a time $t$,
  injected by a homogeneous random driving into a granular gas, in the
  infinite time limit.  Our analytical study starts from a generalized
  Liouville equation and exploits a Molecular Chaos-like assumption. We obtain
  an equation for the generating function of the cumulants $\mu(\lambda)$
  which appears as a generalization of the inelastic Boltzmann equation and
  has a clear physical interpretation.  Reasonable assumptions are used to
  obtain $\mu(\lambda)$ in a closed analytical form. A Legendre transform is
  sufficient to get the large deviation function $\pi_{\infty}(w)$. Our main
  result, apart from an estimate of all the cumulants of ${\cal W}(t)$ at
  large times $t$, is that $\pi_{\infty}(w)$ has no negative branch. This
  immediately results in the failure of the Gallavotti-Cohen Fluctuation
  Relation (GCFR), that in previous studies had been suggested to be valid for
  injected power in driven granular gases. We also present numerical results,
  in order to discuss the finite time behavior of the fluctuations of ${\cal
    W}(t)$. We discover that their probability density function converges
  extremely slowly to its asymptotic scaling form: the third cumulant
  saturates after a characteristic time $\tau$ larger than $\sim 50$ mean free
  times and the higher order cumulants evolve even slower. The asymptotic
  value is in good agreement with our theory.  Remarkably, a numerical check
  of the GCFR is feasible only at small times (at most $\tau/10$), since
  negative events disappear at larger times. At such small times this check
  leads to the misleading conclusion that GCFR is satisfied for
  $\pi_{\infty}(w)$.  We offer an explanation for this remarkable apparent
  verification. In the inelastic Maxwell model, where a better statistics can
  be achieved, we are able to numerically observe the failure of GCFR.
\end{abstract}

\maketitle

\section{Introduction}

In non-equilibrium statistical physics few general results have been
established~\cite{vankampen}. A quick historical survey begins with
the Einstein relation~\cite{einstein} between the velocity
fluctuations at equilibrium of a Brownian particle and the variation
of its velocity under the effect of a small applied force. This
relation between fluctuations and dissipation has been then explored
and fruitfully extended in many ways, first by Onsager~\cite{onsager}, 
and then by Green~\cite{green} and Kubo~\cite{kubo}. 
All those works have been concluded with the establishment of the
Green-Kubo formulae and the fluctuation-dissipation theorem, 
essentially a local formulation of the previous relations. 
Later, a thermodynamical phenomenological description~\cite{degroot} was
formulated, based on the hypothesis of local thermal equilibrium. Within the
latter framework the entropy variations of a macroscopic system evolve
according to the evolution equation
\begin{equation}
{\dd S \over \dd t}= \sigma - 
\int_{\text{system}} \dd V \nabla \cdot {\bf J}_S \ \ .
\end{equation}
The first term $\sigma$ is the irreversible entropy source, which vanishes at
equilibrium, and is strictly positive out of thermal equilibrium. The second
term $\int \nabla \cdot {\bf J}_S$ is related to the external constraints (or
fields), which are necessary to drive the system in a non-equilibrium state.
In a steady state, of course those two terms are equal. Furthermore, as
opposed to $\sigma$, whose origin is purely statistical, ${\bf J}_S$ is the
average of a microscopically well defined quantity.  We shall refer to this
quantity as a mean entropy flux. In many physical systems this entropy flux is
associated to a current, where the conveyed quantities are usually the energy,
the momentum, or the density.

More recently, attempts to reveal the general behavior of large classes of
systems have been performed, focusing on large deviation properties and on
macroscopic quantities. For a long time the study of dissipative systems has
been centered on local quantities, such as structure factors, correlations,
and velocity fields. Nevertheless recent results~\cite{pinton} support the
intuition that in order to observe universal features, it is helpful to study
``global'' quantities.  Averaging is indeed an effective way to bypass
differences arising from microscopic details, keeping however essential
features. Most spectacular and surprising examples are provided
in~\cite{pinton,brey}, where it turns out that three different quantities (the
total magnetization of the two-dimensional XY model near the critical point,
the instantaneous injected power in a turbulent flow, and the energy
fluctuations of a cooling granular gas close to the clustering instability),
suitably rescaled, display the same probability distribution function (pdf).

In addition, we have learnt from equilibrium statistical mechanics that the
most useful informations are contained in the state functions, the free energy
being the main example. Those functions are, in the probabilistic 
interpretation,
large deviation functions (ldf)\footnote{As an example, for a system in the
  canonical ensemble with a free energy $F(T,V,N)$ and a partition function
  $Z=\ee^{-\beta F}$, all the physical features (such phase transitions, {\it
    etc.}) are contained in the free energy per particle $f=F/N$, which in the
  thermodynamic limit is $f=-{1 \over \beta} \lim_{N \to \infty} {\log Z \over
    N}$, which only depends on intensive variables.}. In equilibrium
statistical physics, the large deviation principle enters when considering
microscopic quantities summed over all the degrees of freedom, or integrated
over the whole volume of the system. Thus, large deviations are associated to
the size-extensivity of the macroscopic quantities of interest.  As a matter
of fact in non-equilibrium systems the dynamics, and hence the time, plays a
central role (even in stationary states).  This has led to the introduction
of time extensive quantities, which can be easily obtained integrating a
stationary quantity over a given time interval.  The simplest choice to attempt
for a universal description seems indeed to study the time large deviation
function of a flux integrated over a time $t$.  In this framework the infinite
time limit could be seen as the ``non-equilibrium counterpart'' of the
thermodynamic limit in equilibrium physics.

With this objective Evans, Cohen and Morriss discovered a particular symmetry
property of the ldf of the integrated injected power for a system of
thermostatted sheared hard disks~\cite{evanscohenmorriss}. In order to have a
non-equilibrium steady state, it is necessary to have a competition between at
least two forces acting on the system. In the particular case
of~\cite{evanscohenmorriss} the energy injected by the shearing force is
compensated by an external thermostatting force, created {\it ad hoc}, in such
a way that the total kinetic energy of the system is constant.  This symmetry
property has subsequently been formalized in a theorem by Gallavotti and Cohen
\cite{gallavotticohen}.
The theorem has been proved for deterministic and time reversible dynamical
systems, under the hypothesis of strong chaoticity. In this context the
quantity $\cS(t)$ is the phase space contraction rate integrated over a time
$t$. It turns out that in a large class of dynamical systems the phase space
contraction rate is endowed with the significance of the entropy flux defined
in non-equilibrium thermodynamics. For thermostatted
systems~\cite{thermostats} this is equivalent to the
power injected by the thermostatting force divided by the temperature of the
system~\cite{evanscohenmorriss,gallavotticohen}.  The Gallavotti-Cohen
Fluctuation Relation (GCFR from now on) concerns the large deviation function
of the probability $P(\cS,t)$ of a quantity playing the role of the entropy
flux, defined as
\begin{equation}
\pi_{\infty}(s)=\lim_{t \to \infty} \pi_t(s), \,\,\,\,\,\,\text{with}\,\,\,\,\,
\pi_t(s)=\frac{1}{t} \log P(s t, t)\,.
\end{equation}
Then, the GCFR reads:
\begin{equation}
\label{gcfrldf}
\pi_{\infty}(s)-\pi_{\infty}(-s)=s\,\,\,.
\end{equation}
Similar results were obtained for Markov processes by Kurchan \cite{kurchan}
and by Lebowitz and Spohn \cite{lebowitzspohn}.

After the GCFR has been proposed, several attempts of verification in
experiments and numerical simulations have been proposed, always with
success~\cite{experiments}. In many of these attempts however,
the quantity ${\cS}(t)$
was not the integrated phase space contraction rate, and the original
hypothesis of the theorem were not guaranteed. The phase space contraction rate is
indeed
usually not accessible in experiments, and strong chaoticity (Anosov property)
is not strictly satisfied in most real systems. The choice of the quantity
${\cS}(t)$ usually goes to functionals that are related to equilibrium or
near-equilibrium entropy-like quantities. Typically one expects that the
fluxes defined in non-equilibrium thermodynamics~\cite{degroot} are the
macroscopic analogous of the phase space contraction rate, and this is
definitely true in specific thermostatted models~\cite{thermostats}. The rate
of heat transport from the thermostat, or equivalently the power injected by
the thermostat into the system, divided by the temperature of the thermostat,
is the natural candidate. In order to check for possible extensions of the
GCFR for systems without time-reversal invariance the study of the pdf of the
injected power seems the simplest choice. Such a program has been initiated by
Auma\^\i tre {\it et al.}  \cite{aumaitrefauvemcnamarapoggi}, who measured the
injected power distribution for several microscopically irreversible systems.
Within the available accuracy, they concluded that the GCFR holds but
importantly pointed out the possible relevance of spurious effects at large
times where the necessarily limited data gathered may not allow to sample the
regions where GCFR violations could exist.  An analytical approach was also
proposed by Farago \cite{farago}, who analytically computed the ldf of a
particular model without time-reversal symmetry, 
showing that GCFR does not hold, at
least for this particular model. Nevertheless a recent
experiment~\cite{feitosamenon} observed the validity of the GCFR
measuring the integrated injected power rendered dimensionless by an
appropriate energy scale.

In a recent study, we have shown that in general the GCFR is not satisfied 
for the integrated injected power in
homogeneously driven granular gases~\cite{letter2} and that the results of the
experiment described in~\cite{feitosamenon} 
were insufficient to claim its
verification~\cite{letter1}. In this paper we discuss the derivation of an
equation for the large deviation function for injected power in homogeneously
driven granular gases, obtaining its solution under a series of reasonable
approximations. The Gallavotti-Cohen Fluctuation Relation cannot be satisfied
in this model because the injected power large deviation has no negative
tail. We also provide details from numerical simulations of the model, showing
how the GCFR may seem to be satisfied at small times 
[i.e. considering $\pi_t(s)$ instead of $\pi_\infty(s)$]
and small values of $s$. In
experiments and simulations true large deviations are argued to be unreachable.

In our approach, the granular gas is modeled as a dilute system of
inelastic hard spheres. The heating mechanism that allows the gas to reach a
steady-state is the widely used stochastic thermostat~\cite{kicks}. It differs
from the experimental setup of Ref.~\cite{feitosamenon} in that the energy
injection is fully uniform (each particle is subjected to a random force)
instead of being boundary driven. The model under study is however relevant
for other two dimensional experimental situations, see e.g. \cite{prevost}.
Our goal is to compute, analytically and numerically, the distribution of the
power injected into the gas by those random kicks.

In section \ref{sec:IHS} 
the inelastic Hard Spheres model is presented and the details of
the derivation of the large deviation function of the injected power are
presented. Section \ref{sec:Maxwell} 
repeats the procedure for the closely related and much
more tractable inelastic Maxwell Model (practical motivations will become
clear later). Numerical results are provided in section \ref{sec:numerics}. 
We then conclude
the paper (Section \ref{sec:conclusion}) with a discussion of our results.

\section{The Inelastic Hard Spheres Model}
\label{sec:IHS}

\subsection{The Model}
We consider a granular gas made up of $N$ identical spherical particles
undergoing dissipative collisions.  The collisions between two particles $i$
and $j$ conserve the total momentum and reduce the normal component of the
relative velocity, i.e. $(\mathbf{v}_i^*-\mathbf{v}_j^*)\cdot \bsig=-\alpha
(\mathbf{v}_i-\mathbf{v}_j)\cdot \bsig$, where the stars denote the
post-collisional velocities and $\bsig$ the direction joining the centers of
the colliding particles. Energy injection, achieved by means of random forces
(kicks)
acting independently on each particle, drives the gas into a non-equilibrium
steady state. The equation of motion governing the dynamics of each particle
is therefore:
\begin{equation}
{\ddr \bv_i \over \ddr t} = \bF_i^{\text{coll}} + \bF_i^{\text{th}}, 
\end{equation}
where $\bF_i^{\text{coll}}$ is the force due to collisions and
$\bF_i^{\text{th}}$ is a Gaussian white noise (i.e. $\langle
F_{i\gamma}^{\text{th}}(t)F_{j\delta}^{\text{th}}(t')\rangle=2 \Gamma
\delta_{ij} \delta_{\gamma\delta}\delta(t-t')$, where the angular brackets
denote statistical average; the subscripts $i$ and $j$ are used to refer to
the particles, while $\gamma$ and $\delta$ denote the Euclidean components of
the random force).  This model is one of the most studied in granular gas
theory and reproduces many qualitative features of real driven inelastic
gases~\cite{kicks,vannoijeernst}.  After a few collisions per particle, the
system attains a non-equilibrium stationary state. Furthermore, this state is
homogeneous and does not develop spatial 
inhomogeneities, in contrast to what happens for
example in the freely cooling state of a granular gas. From the equations of
motion it is possible to derive the Boltzmann equation governing the evolution
of the one-particle velocity distribution function $f$ which,
since the
system is homogeneous, will not depend on the positions of the
particles. If, in addition, we resort to the molecular chaos assumption, this
Boltzmann equation reads~\cite{vannoijeernst}:
\begin{equation}
\label{boltz}
\partial_t f(\bv_1, t) = J[f,f] +\Gamma \Delta_{\bv_1} f(\bv_1)\,\,,
\end{equation}   
where the Laplace operator $\Delta_{\bv}\equiv (\p / \p_{\bv})^2$ is a
diffusion term in velocity space characterizing the effect of the random
force, while $J[f,f]$ is the collision integral, which takes into account the
inelasticity of the collisions:
\begin{equation}
\label{collintboltz}
  J[f,f]={1 \over \ell} \int \ddr \bv_2 \int' 
    \ddr \bsig (\bv_{12} \cdot \bsig) \left({1 \over
    \alpha^2}f(\bv_1^{**}) f(\bv_2^{**}) - f(\bv_1)f(\bv_2) 
\right) \,\,.
\end{equation}   
In the above expression of the collision integral the notation $\bv_{12}$
denotes the relative velocity between particles 1 and 2, while the two stars
superscript (i.e. $\bv^{**}$) denotes the pre-collisional velocity of a
particle having velocity $\bv$. Moreover the length $\ell=(n
\sigma^{d-1})^{-1}$ , where $n$ is the particle density, $\sigma$ the diameter
of the particles and $d$ the space dimension, is proportional to the mean free
path. Finally, the primed integral $\int'$ will be used as a short-hand
notation to denote $\int \Theta(\bv_{12} \cdot \bsig)$, where $\Theta$ is the
Heaviside step function. The granular temperature of the system is defined as
the mean kinetic energy per degree of freedom, $T_g=\langle v^2 \rangle /d$.
Such a definition for the temperature is purely kinetic, and can {\it a
  priori} not be interpreted as a thermodynamical temperature.

\subsection{The velocity distribution function}
In this section we briefly review the known results concerning the solution of
the Boltzmann equation (\ref{boltz}). The stationary solution of this equation
has extensively been investigated in the last years, but an exact solution is
still missing.  Nevertheless a general method is to look for solutions in the
form of a Gaussian distribution multiplied by a series of Sonine polynomials:
\begin{equation}
\label{stationarypdf}
  f_{st}(\bv)=e^{-{v^2 \over 2 T_g }} 
  \left(1+ \sum_{p=1}^{\infty} a_p
    S_p\left({v^2 \over 2 T_g} \right) \right)\,\,.
\end{equation}
The expression of the first three Sonine polynomials is:
\begin{gather}
  S_0(x)=1   \notag \\
  S_1(x)=-x + {1 \over 2} d  \\
  S_2(x)={1 \over 2} x^2 -{1 \over 2} (d+2)x + {1 \over 8} d(d+2) \,\,.\notag 
\end{gather}
Moreover the coefficients $a_p$ are found to be proportional to the averaged
polynomial of order $p$:
\begin{equation}
\label{coefficients}
a_p=A_p \left\langle S_p \left({v^2 \over 2 T_g} \right) \right\rangle\,\,,
\end{equation}
where $A_p$ is a constant. From this observation one directly obtains that the
first coefficient $a_1$ vanishes by definition of the temperature.  A first
approximation for the velocity pdf is therefore to truncate the expansion at
second order ($p=2$).  An approximated expression for the coefficient
$a_2$ has been found as a function of the restitution coefficient $\alpha$ and
the dimension $d$ \cite{vannoijeernst, coppex, montanerosantos}.  It must be
noted that this approximation is only valid for not too large velocities,
since the tails of the pdf have been shown~\cite{vannoijeernst} to be
overpopulated with respect to the Gaussian distribution. It is
known~\cite{vannoijeernst} that at high energies $\log f(\bv) \sim
-(v/v_c)^{3/2}$ with a threshold velocity $v_c$ that diverges when the
dimension $d$ goes to infinity. This means that at high dimensions the
distribution is almost a Gaussian with small Sonine corrections. All the above
results have been confirmed by numerical simulations, in particular through
Molecular Dynamics (MD) and Direct Simulation Monte Carlo (DSMC) \cite{bird}
methods. Those two numerical methods, although very different, show a
surprisingly good agreement. This points out the validity of 
the molecular
chaos assumption and thus the relevance of the DSMC method, which is
particularly well adapted to simulate the dynamics of a homogeneous dilute
gas.

\subsection{Equation for the injected power}

In this section, we will show how to obtain a kinetic equation able to
describe the behavior of the pdf of the integrated injected power at large
times. The latter quantity is the total work ${\cal W}$ provided by the
thermostat over a time interval $[0,t]$:
\begin{equation}
{\cal W}(t) = \int_0^t \dd t \sum_i \bF^{th}_i \cdot \bv_i\,.
\end{equation} 
Our interest goes to the distribution of ${\cal W}(t)$, denoted by $P({\cal
  W},t)$, and to its associated large deviation function $\pi_{\infty}(w)$
defined for the reduced variable $w={\cal W}/t$ (${\cal W}(t)$ being extensive
in time):
\begin{equation}
\pi_{\infty}(w)=\lim_{t \to \infty} \pi_t(w) \,\,\,,
\pi_t(w) =\frac{1}{t}\ln P({\cal W}=wt,t)\,.
\end{equation}
We introduce $\rho(\Gamma_N,{\cal W},t)$, the probability that the system is in
state $\Gamma_N$ at time $t$ with ${\cal W}(t)={\cal W}$. The function we want
to calculate is
\begin{equation}
P({\cal W},t)=\int\dd\Gamma_N \rho(\Gamma_N,{\cal W},t) \ \ .
\end{equation}
We shall focus on the generating function of the phase space density
\begin{equation}
\hat{\rho}(\Gamma_N,\lambda,t)=\int\dd {\cal W}\ee^{-\lambda 
{\cal W}}\rho(\Gamma_N,{\cal W},t)
\end{equation}
and on the large deviation function of
\begin{equation}
\hat{P}(\lambda,t)=\int\dd
{\cal W}\ee^{-\lambda {\cal W}}P({\cal W},t)=
\int\dd\Gamma_N\hat{\rho}(\Gamma_N,\lambda,t) \ \ ,
\end{equation}
which we define as
\begin{equation}
\mu(\lambda)=\lim_{t \to \infty} \frac{1}{t}\ln \hat{P}(\lambda,t)\,\,.
\end{equation}
Note that $\mu(\lambda)$ is the generating function of the cumulants of ${\cal
  W}$, namely
\begin{equation}
\label{cumulants}
\lim_{t \to \infty} \frac{\langle {\cal W}^{n}\rangle_c}{t}=
(-1)^n \left.\frac {\dd^n\mu(\lambda)}{\dd\lambda^n}\right|_{\lambda=0} \ \ .
\end{equation}
Moreover $\pi_{\infty}(w)$ can be obtained from $\mu(\lambda)$ by means of a
Legendre transform, i.e. $\pi_{\infty}(w)=\mu(\lambda_*)+\lambda_*w$ with
$\lambda_*$ such that $\mu'(\lambda_*)=-w$.

The observable ${\cal W}$ is non-stationary and we are interested in the
evolution equation for the extended phase space density
$\rho(\Gamma_N,\cW,t)$. It varies in time under the combined effect of the
inelastic collisions (which do not trigger any change in $\cW$) and of the
random kicks:
\begin{equation}
\p_t\rho=\p_t\rho\Big|_{\text{collisions}}+\p_t\rho\Big|_{\text{kicks}}
\end{equation}
Instantaneous collisions do not modify the value of $\cW$, only the random
kicks will affect it. In between two successive collisions, the amount of
energy provided by the thermostat on each particle is $\Delta
\cW=(v(t_{i+1}^-)^2-v(t_i^+)^2)/2$, where $t_i$ denotes the time of the $i-$th
collision and the $-$ ($+$) superscript simply denotes that the value of the
velocity at a collision time has to be taken before (after) the collision has
taken place. 

In order to characterize the role of the thermostat on the generating function
$\hat \rho$ it is useful to discretize the problem in velocity space,
considering that the effect of the random kicks on one particle is a
continuous time random walk of its velocity on a $d$ dimensional lattice of
step $a$, with hopping rate $\gamma$. It then appears that the observable
$\cW$ is Markovian, and the master equation governing the evolution of the
probability of being at site $\bn$ (which represents the discretized velocity
of one particle) at time $t$, with the value ${\cal W}(t)=\bn(t)^2/2$ reads:
\begin{equation}
\p_t P(\bn,{\cal W},t)=\gamma \sum_{\ba} \left( P(\bn+\ba ,{\cal W}+
\bn \cdot \ba+a^2/2,t)+ P(\bn-\ba,{\cal W}-\bn\cdot\ba+a^2/2,t) \right)
-2d \gamma P(\bn,{\cal W},t)\,\,.
\end{equation} 
Here $\{\ba\}$ is a set of $d$ orthogonal vectors defining the lattice.
This yields, in terms of the generating function
$\hat{P}(\bn,\lambda',t)=\int\dd {\cal W}\ee^{-\lambda' {\cal W}}P(\bn,{\cal
  W},t)$,
\begin{equation}
\p_t \hat{P}(\bn,\lambda',t)=\gamma \sum_{\ba}
\left( \hat{P}(\bn+\ba,\lambda',t)\ee^{\lambda' \bn\cdot\ba+
\lambda' a^2/2}+
\hat{P}(\bn-\ba,\lambda',t)\ee^{-\lambda' \bn\cdot\ba+\lambda'
a^2/2} \right)-2d\gamma \hat{P}(\bn,\lambda',t).
\end{equation}
Next, it is possible to expand the above expression in powers of $a$ at
order $a^2$ and consider the continuum limit $a \to 0$ with the scaling
variables $\bv =\bn \,a$, $\Gamma=\gamma a^2$ and $\lambda=\lambda'/ a$ fixed.
Finally this yields, considering that the thermostat acts independently on
each particle,
\begin{equation}
\p_t\hat{\rho}\Big|_{\text{kicks}}=
\sum_i\Big[ \Gamma(\Delta_{\bv_i}+2\lambda \Gamma
\bv_i \cdot \p_{\bv_i}+\Gamma(d\lambda+\lambda^2 v_i^2)\Big]\hat{\rho}
\end{equation}
This additional piece is linear in $\hat{\rho}$, just as the collision part.
The large time behavior of $\hat{\rho}$ is thus
governed by the largest eigenvalue
$\mu(\lambda)$ of the evolution operator of $\hat{\rho}$. In the large time
limit, we expect that
\begin{equation}
\hat{\rho}(\Gamma_{N},\lambda,t)\simeq C(\lambda) \ee^{\mu(\lambda) t}\tilde
{\rho}(\Gamma_N,\lambda),
\end{equation}
where $C(\lambda)=1/\int \dd \Gamma_N \tilde{\rho}(\Gamma_N, \lambda)$, and
$\tilde{\rho}(\Gamma_N,\lambda)$ is the eigenfunction associated to $\mu$,
which for convenience has been taken normalized to unity (in order that
$C(\lambda)\neq 0$ the initial state must have a nonzero projection onto this
eigenfunction).  We then introduce
\begin{equation}\hat{f}^{(1)}(\bv,\lambda,t)=\int\dd\Gamma_{N-1}\hat{\rho},
\end{equation}
where $\int\dd\Gamma_{N-1}$ means an integration over $N-1$ particle
coordinates. This function obeys
\begin{equation}\label{base}
\p_t\hat{f}^{(1)}(\bv,\lambda,t)=\Gamma\Delta_{\bv}\hat{f}^{(1)}
+2\lambda \Gamma
\p_{\bv} \cdot \bv \hat{f}^{(1)}
+\Gamma(\lambda^2 v^2-d \lambda)\hat{f}^{(1)}+\hat{J}
\end{equation}
with $\hat{J}=\int\dd {\cal W}\ee^{-\lambda {\cal W}}J$ the Laplace transform
of the collision integral $J$ in which the usual 
velocity distributions $f^{(1)}(\bv_1,t)$ and $f^{(2)}(\bv_1,\bv_2,t)$
are replaced by ${f}^{(1)}(\bv,{\cal W},t)$ and
$f^{(2)}(\bv_1,\bv_2,{\cal W},t)$ involving $\cal W$ as well as the velocities.
Quite unexpectedly the above equation has a
clear physical interpretation: consider a many particle system where a
noise of strength $\Gamma$ and a viscous friction-like force $\bF=-2 \lambda
\Gamma \bv$ act independently on each particle, and where the particles
interact by inelastic collisions. Consider then that the particles
annihilate/branch (depending on the sign of $\lambda$) at constant rate $d
\lambda \Gamma$, and branch with a rate proportional to $\lambda^2 v^2
\Gamma$. Then, the equation governing the evolution of the one particle
velocity distribution of such a system is exactly the equation (\ref{base}),
where $\lambda$ is a parameter tuning the strength of the external fields. In
spite of there being no {\it a priori} reason for that, $\tilde \rho$, as well
as $\tilde f^{(1)}=\int \dd \Gamma_{N-1} {\tilde \rho}$, can consequently be
interpreted as probability density functions.

Since 
$\hat{\rho}(\Gamma_{N},\lambda,t)\simeq C(\lambda) \ee^{\mu(\lambda) t}\tilde
{\rho}(\Gamma_N,\lambda)$,
the one and two-point functions ${f}^{(1)}(\bv,{\cal W},t)$ and
$f^{(2)}(\bv_1,\bv_2,{\cal W},t)$ that enter the expression of $J$ are
expected to verify in $\lambda-$space, at large times,
\begin{equation}
\hat{f}^{(1)}(\bv_1,\lambda,t)= C(\lambda) \ee^{\mu t}
\tilde{f}^{(1)}(\bv_1,\lambda),
\end{equation}
and
\begin{equation}
\hat{f}^{(2)}(\bv_1,\bv_2,\lambda,t)=C(\lambda) \ee^{\mu t}
\tilde{f}^{(2)}(\bv_1,\bv_2,\lambda),
\end{equation}
where both $\ft^{(1)}=\int \dd \Gamma_{N-1} {\tilde \rho}$ 
and $\ft^{(2)}=\int \dd \Gamma_{N-2} {\tilde \rho}$ are 
normalized to unity.  We perform
the following molecular-chaos-like assumption:
\begin{equation}\label{truncation}
\tilde{f}^{(2)}(\bv_1,\bv_2,\lambda)\simeq\tilde{f}^{(1)}(\bv_1,\lambda) 
\tilde{f}^{(1)}(\bv_2,\lambda)
\end{equation}
which does have a definite physical interpretation in the language of the
inelastic hard-spheres with fictitious dynamics (viscous friction, velocity
dependent branching/annihilation) described in the above paragraph.  Then we
get that
\begin{equation}\label{lambdaBoltzmann}\begin{split}
    \mu\tilde{f}(\bv,\lambda,t)=&\Gamma\Delta_{\bv}\tilde{f}+2\lambda \Gamma
    \bv\cdot\p_{\bv}\tilde{f}+\Gamma(d\lambda+\lambda^2 v^2)\tilde{f}\\&+
    \frac{1}{\ell}\int_{\bv_{12}\cdot\bsig>0}\dd \bv_2 \dd\bsig
    \bv_{12}\cdot\bsig
    \left[\alpha^{-2}\tilde{f}(\bv_1^{**},\lambda)\tilde{f}(\bv_2^{**},\lambda)
      -\tilde{f}(\bv_1,\lambda)\tilde{f}(\bv_2,\lambda)\right]
\end{split}
\end{equation}
where we have now omitted the superscript $(1)$ denoting the one-point
function. Note that the truncation (\ref{truncation}) respects two physical
requirements. First, it does conserve ${\cal W}$ throughout a collision.
This can be observed rewriting the product of the two one-point functions as
\begin{equation}
\ft \left(\bv_1^{(**)},\lambda \right) \ft \left(\bv_2^{(**)}, \lambda \right)=
\int \dd \cW \, \dd \cW_1 \, \dd \cW_2 \, \delta(\cW_1 +\cW_2 - \cW) 
\ee^{-\lambda \cW} g \left(\bv_1^{(**)},\cW_1 \right) g \left(\bv_2^{(**)}, 
\cW_2 \right)\,\,,
\end{equation}
where $g(\bv, \cW)=\int \dd \Gamma_{N-1} \, \rho(\Gamma_N, \cW)$. This leads
to the simple remark that if the particle 1 carries a fraction $\cW_1$ of the
integrated injected power before a collision, and the particle 2 carries a
fraction $\cW_2$, then the global amount $\cW=\cW_1 + \cW_2$ is preserved in a
collision. Second, the $\lambda=0$ limiting case yields the usual Boltzmann
equation, since in this case a stationary solution exists, and hence $\mu
(\lambda=0) =0$. The boundary condition to the evolution equation above is
thus:
\begin{equation}
\tilde{f}(\bv,\lambda=0)=f_{st}(\bv)
\end{equation}
with $f_{st}(\bv)$ the stationary velocity pdf (cf. Eq.(\ref{stationarypdf})).


\subsection{The limits for large velocity  and for $\lambda \to \infty$.}

Equation (\ref{lambdaBoltzmann}) can be written in the form:
\begin{equation}
\mu\tilde{f}(\bv_1,\lambda)=\Gamma \ee^{-{\lambda v^2 \over 2}} \Delta_{\bv_1}
  \left(\ee^{\lambda v_1^2 \over 2}\tilde{f}(\bv_1,\lambda)\right) 
+{\hat J}(\ft,\ft)\,\,.
\end{equation}
Defining a new function $F(\bv,\lambda)=\exp{\left( \lambda v^2 \over 2
  \right)}\tilde{f}(\bv,\lambda)$ the above equation reads:
\begin{equation}
\label{eqforF}
\mu(\lambda) F(\bv_1,\lambda)=\Delta_{\bv_1} F(\bv_1,\lambda) 
+ J_{\lambda}[F,F]\,\,,
\end{equation}
where
\begin{equation}
J_{\lambda}[F,F]=\int \ddr \bv_2 \int' \ddr \bsig \,(\bv_{12} \cdot \bsig)\,
\ee^{-{\lambda v_2^2 \over 2}}\left({1 \over \alpha^2}
\ee^{- {\lambda \over 4}\left({1 \over \alpha^2}-1 \right) 
(\bv_{12}\cdot \bsig)^2} F(\bv_1^{**}, \lambda) F(\bv_2^{**}, \lambda)
-F(\bv_1, \lambda) F(\bv_2, \lambda) \right)
\end{equation}
is the collision integral. Repeating the arguments by van Noije and Ernst
\cite{vannoijeernst} we shall now compute the behavior of the tails of $\ft$.
In the large velocities limit one has $v_{12} \sim v_1$, and for $\lambda >0$
the gain term of the collision integral can be neglected, since the argument
of its exponential prefactor is strictly negative for inelastic collisions
($\alpha <1$). In this limit the solution of eq.(\ref{eqforF}) is of the form:
\begin{equation}
\log \ft(\bv,\lambda)=-\lambda {v^2 \over 2}- {2 \over 3} \sqrt{\beta_1 \over
  \ell \Gamma} v^{3/2} - {\mu \over \Gamma} \sqrt{\ell \Gamma \over \beta_1}
  v^{1/2} +  o(v^{1/2})\,\,,
\end{equation}
where $\beta_1=\pi^{(d-1)/2}/\Gamma((d+1)/2)$ comes from an angular
integration. This result is important since it shows that the high velocity
tails of the function $\ft(\bv,\lambda)$ are Gaussian.  For $\lambda=0$ the
known result concerning the non-Gaussian tails of the one-particle velocity
pdf~\cite{vannoijeernst} is recovered.

Moreover, in the limit $\lambda \to \infty$ the gain term of the collision
integral can be neglected. Thus, the resulting equation can be interpreted as
a Boltzmann equation for a system of particles subjected to a random force,
interacting with an ``effective potential'' which makes the collision kernel
$\propto (\bv_{12} \cdot \bsig ) \exp{-\lambda v_2^2/2}$, and annihilating
when a collision occurs.  The Boltzmann equation of this system reads:
\begin{equation}
\partial_t F=\Delta_{\bv} F -\int \ddr\bv_2 \int' \ddr \bsig (\bv_{12}
\cdot \bsig ) e^{-{\lambda
    v_2^2 \over 2}} F(\bv_1)F(\bv_2)\,\,.
\end{equation}
This immediately shows that $\mu(\lambda)$ for high values of $\lambda$ is
always negative, since the density of particles (given by the integral of $F$)
is clearly decreasing. The consequence of
this observation, together with the fact that $\mu(\lambda)$ must be always
convex and is zero in $\lambda=0$, is that $\mu'(\lambda)<0$ for any value of
$\lambda$ . For negative $w$, there is therefore no possible 
$\lambda_*$ satisfying $\mu'(\lambda_*)=-w$: the immediate
consequence is that $\pi_{\infty}(w)$,
i.e. the Legendre transform of $\mu(\lambda)$, is not defined for negative
$w$. As announced in the introduction, this is a key point in invalidating 
the GCFR.

\subsection{Energy balance}\label{EnergyBalance}

Before going further in the analytical calculation of $\pi_{\infty}(w)$, we
provide a complementary argument for the absence of negative tail.
The reasoning takes as starting point the fact that
the energy variations in a time $t$ are given by the
difference of the injected power integrated over a time $t$, ${\cal W}(t)$, and
the energy dissipated through collisions over the same time interval ${\cal
  D}(t)$. Hence the fluctuating total kinetic energy $E(t)=\sum_i \bv_i^2/2$
varies according to
\begin{equation}
\Delta E(t)=E(t)-E(0)={\cal W}(t)-{\cal D}(t)\,\,,
\end{equation}
where ${\cal D} \ge 0$.  It is important to note that the above equation is
very general, and applies to many dissipative systems where an external
driving supplies the energy lost by some dissipation mechanism in order to
reach a non-equilibrium steady state. Let us define the probability ${\cal
  P}(z)$ of having $\Delta E(t)=z$. If the total work $\cW (t)$ is sampled
starting from a fixed initial condition $E(0)=const$, then it is easy to show
that the left tail of the distribution ${\cal P}$ is bounded by $-E(0)$
(since $E(t) \ge 0$). This
implies that the large deviation function associated to the stationary
quantity $\Delta E(t)$ has no negative contributions, as well as the large
deviation function $\pi_{\infty}$ of the total work $\cW=\cD+ \Delta E$, since
it is the sum of two quantities with no large negative events in the long times
limit. On the contrary, if the total work distribution is obtained sampling
over segments of a unique trajectory, the above argument does not hold
anymore, since
both $E(t)$ and $E(0)$ are fluctuating quantities. Thus, the pdf $\cP(z)$ is
no more bounded, but, for times much longer than the characteristic energy
correlation time, it is symmetric with respect to the $z=0$ axis.  Our purpose
in this section is to give some phenomenological arguments to understand under
which conditions on $\cP (z)$ it is likely that $\cW$ and $\cD$ have the same
large deviations function. Since $\Delta E$ is not a time-extensive variable,
in principle it does not have large deviations. 
Nevertheless a large fluctuation of
$\Delta E$ may affect the behavior of the time-extensive quantities $\cW$ or
$\cD$. Hence we will focus our attention to the tails of $\Delta E$, which we
will suppose to have a stretched exponential behavior with an exponent
$\delta$ (i.e. $\cP (z) \sim \exp-|z|^{\delta}$ when $z \to \pm \infty$). It
is useful then to note that
\begin{equation}
\label{ldfenergy}
\zeta(\epsilon)=\lim_{t \to \infty} \frac{1}{t} \log \cP (\Delta E=\epsilon
t)=
\begin{cases}
0\,\,, & \text{if $\delta < 1$}  \\
- |\epsilon | \,\,, & \text{if $\delta = 1$} \\
- \infty \,\,, & \text{if $\delta > 1$ (and $\epsilon \ne 0$)}.
\end{cases}
\end{equation} 
We therefore observe that the exponential distribution ($\delta=1$) plays a
limiting role for the function $\zeta$. Besides, if one looks at what happens
in Laplace space (i.e. to the large deviation function of the Laplace
transform of $\cP$), the infinite time limit does not depend anymore on 
the particular value of the exponent $\delta$:
\begin{equation}
\vartheta(\lambda)=\lim_{t \to \infty}
\frac{1}{t} \log \langle \ee^{\lambda \Delta E} \rangle = 0\,\,.
\end{equation}
Now a natural question arise: is it possible to recover the function $\zeta$
knowing $\vartheta$ ? Usually the Legendre transform is the link
between those two functions. It is easy to see that the Legendre
transform of $\vartheta=0$ gives $\zeta=-\infty$, independently of $\delta$.
As we will show, this ``failure'' of the Legendre transform is related with
analyticity breaking of the function $\vartheta$. If $\delta >1$ the Laplace
transform of $\cP$ is analytical in the whole complex plane of $\lambda$. On
the contrary, if $\delta=1$, the Laplace transform of $\cP$ is very likely to
presents singularities, as well as analytical cuts in the complex plane of
$\lambda$. The Legendre transform as an inversion formula of the Laplace
transform is obtained using the saddle point method when integrating the
inverse Laplace integral on a straight line parallel to the imaginary axis of
the complex $\lambda$ plane. It is seen indeed that if $\langle \exp \lambda
\Delta E \rangle$ presents singularities or cuts, then the Laplace transform
cannot be inverted anymore.

Returning to our original problem, it now seems reasonable to argue that in a
general way the large deviations functions associated to the Laplace transform
of the pdf of $\cW$ and $\cD$ are the same (i.e.  $\lim \langle \exp (-\lambda
\cW) \rangle /t = \lim \langle \exp (-\lambda \cD) \rangle /t$). However, if
$\cP (\Delta E)$ has exponential tails, non-analyticities may arise in Laplace
space, and the large deviations functions of $\cW$ and $\cD$ could be
different. Besides, if $\cP ( \Delta E)$ has tails decreasing faster then the
exponential, the large deviations functions of $\cW$ and $\cD$ are the same.

This scenario gives then simple hints for the knowledge of the large deviation
functions of two time-extensive quantities which differs only by a stationary
term. It turns out that if this stationary term is distributed with
exponential tails, its fluctuations may affect the large deviation functions
of the time-extensive quantities. This is clearly already seen in eq.
(\ref{ldfenergy}), which shows that in this case fluctuations of order $t$ are
possible. This clearly does not happen if the tails of the pdf of the
stationary term decrease faster than any exponential. In this case
fluctuations of order $t$ vanish in the infinite time limit, and the two large
deviation functions are equal. Moreover, this scenario is perfectly in line
with the explicit results obtained by Farago in several simple models
\cite{farago, farago2}, and by Van Zon and Cohen in
\cite{vanzoncohen}. In all those solvable cases the system considered is a
one-particle system, with a very low number of degrees of freedom. Hence, if
the velocity of the particle has a Gaussian distribution, the energy pdf has
exponential tails, and in principle the distribution of $\cW $ and $\cD$ can
be different. Nevertheless, if the system under investigation is a many
particle system, as it is the case here, the energy pdf is distributed
following the gamma distribution \cite{feller}:
\begin{equation}
P(E) \sim E^{{N \over 2} -1} \exp (- E)\,\,
\end{equation}
where $N$ is the number of degrees of freedom of the system. Moreover, in the
thermodynamic limit (i.e. when $N \to \infty$) the above expression of $P(E)$
approaches the Gaussian distribution. Thus, taking the thermodynamic limit
before that of large times, it follows from the above discussion that for
large systems (i.e. in the thermodynamic limit) the total work $\cW$ and the
dissipated energy $\cD$ have the same large deviation function. This is in
line with the observations of the previous subsection, which qualitatively
predict the absence of a negative tail for $P(\cW)$.

\subsection{The cumulants}

\begin{figure}
\begin{minipage}[t]{.46\linewidth}
  \includegraphics[width=1 \textwidth]{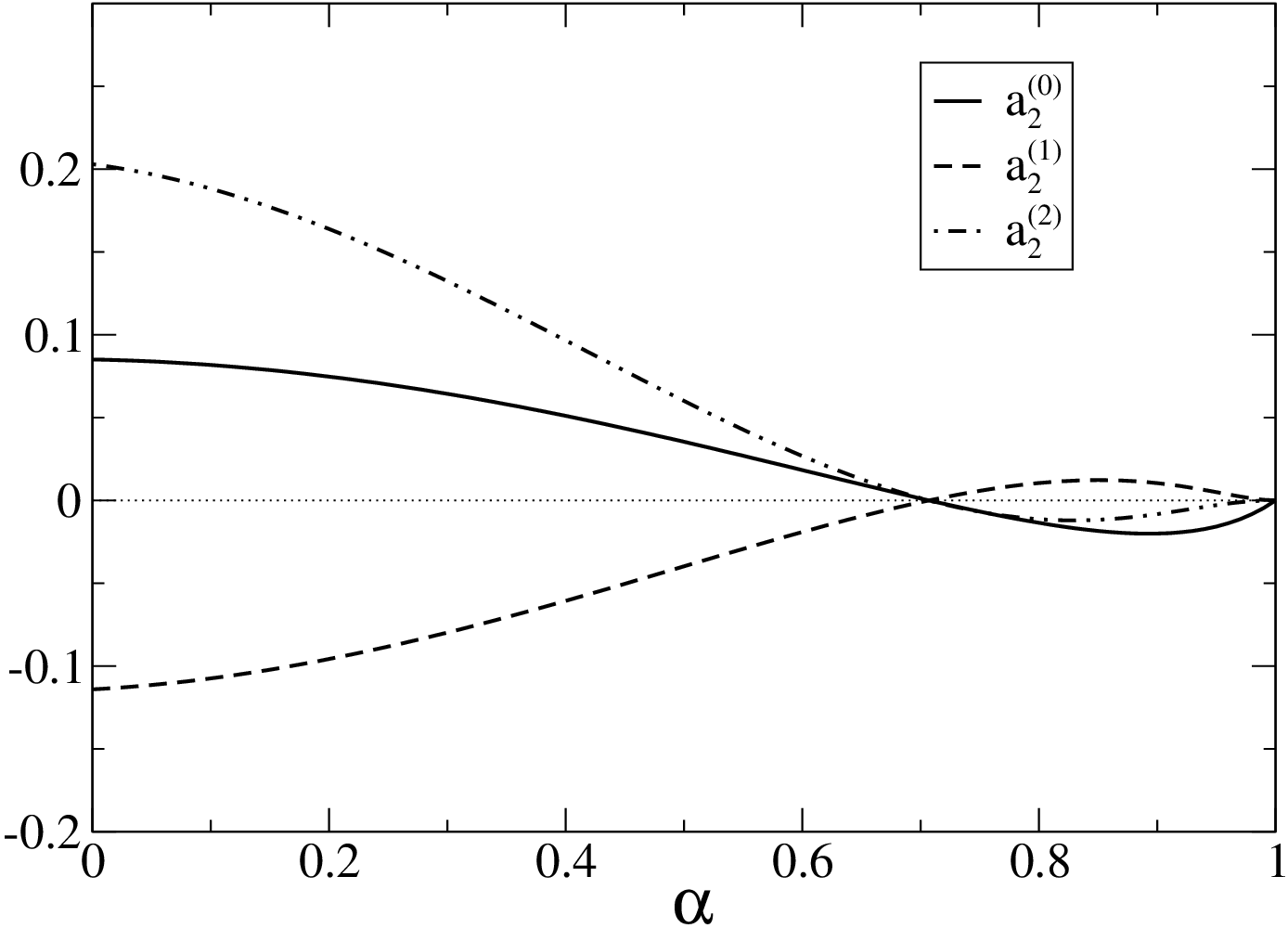}
\caption{\label{a2plot}
$a_2^{(0)}$, $a_2^{(1)}$ and $a_2^{(2)}$ versus $\alpha$ for 
inelastic hard spheres driven by random forces in $d=2$.}
\end{minipage}
\hfill
\begin{minipage}[t]{.46\linewidth}
  \includegraphics[width=1 \textwidth]{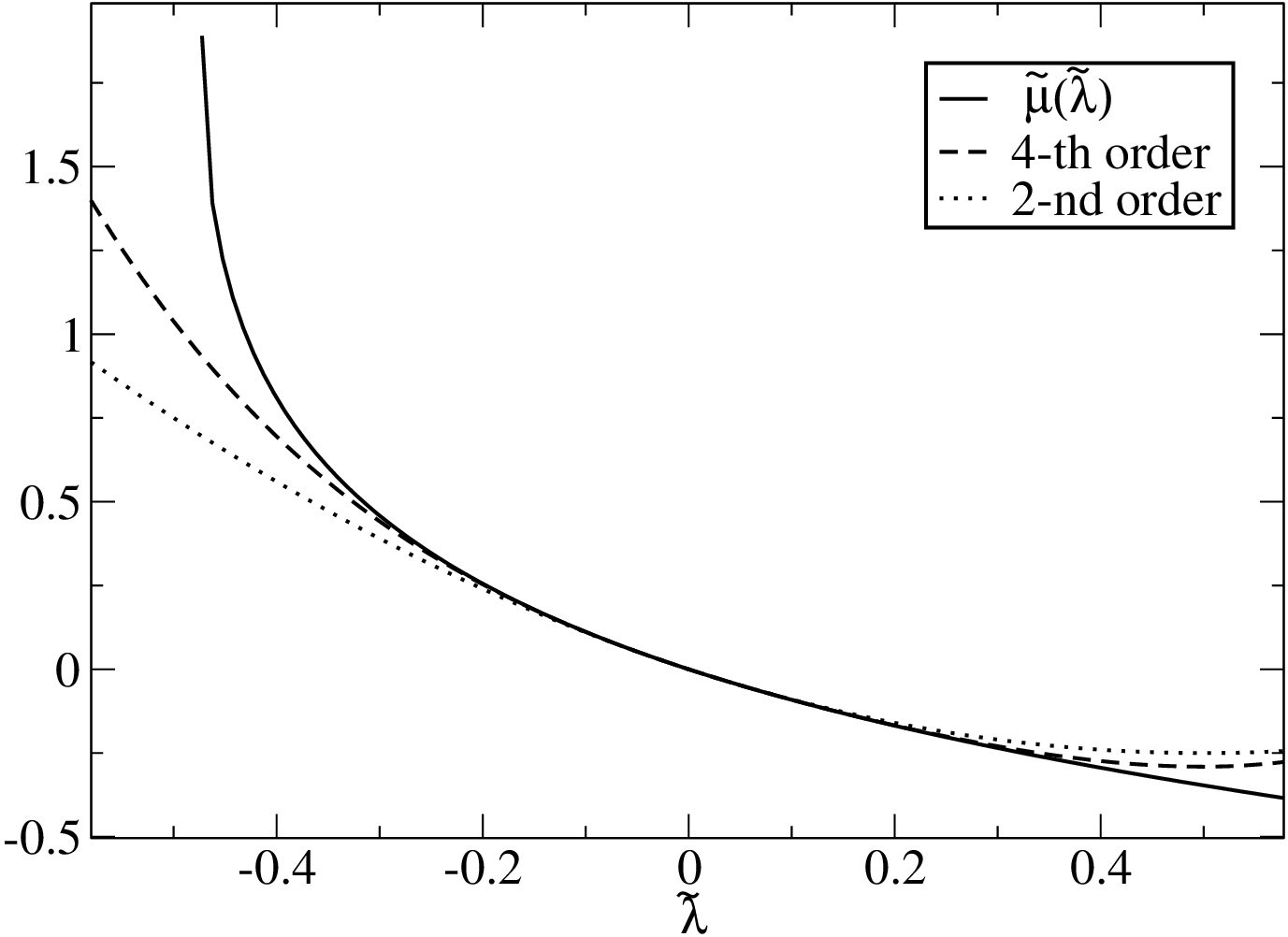}
\caption{\label{plotmunoappr}The solid line shows $\mut$ in the
  limit $d \to \infty$ for inelastic hard spheres. 
  The dashed line is $\mut$ at fourth order in $\lambt$
  from (\ref{solsys1}) for $d=2$ and $\alpha=0.5$. Finally the dotted line
  shows the same quantity calculated with a truncation at second order in
  $\lambda$, which would satisfy the GCFR. }
\end{minipage}
\end{figure}

In this section we find an approximated expression of $\mu(\lambda)$ solving a
system of equations obtained projecting (\ref{lambdaBoltzmann}) on the first
velocity moments.  First we shall define a dimensionless velocity ${\bf
  c}={\bf v}/v_0(\lambda)$, where $v_0(\lambda)$ plays the role of a thermal
velocity:
\begin{equation}
\label{v02lambda}
v_0^2(\lambda)=2 T (\lambda) = {2 \over d} \int \textrm{d}{\bf v}  \, v^2 \, 
{\tilde f}(\bv,\lambda)\,.
\end{equation}
Then, defining the function $f(\bc,\lambda)=v_0(\lambda)^d{\tilde f}(\bv,
\lambda)$, and its related moments of order $n$
\begin{equation}
\label{remoments}
m_n (\lambda)= \int \textrm{d}{\bf c} \, c^n f(\bc,\lambda) \,\,\,,
\end{equation}
one obtains the following recursion relation:
\begin{equation}
\label{moments}
(\mu + \Gamma(2n+d) \lambda)m_n=
{\Gamma \over v_0^2} n(n+d-2)m_{n-2}+\Gamma \lambda^2 v_0^2 
m_{n+2}- v_0   \nu_n,
\end{equation}
where 
\begin{equation}
\nu_n=-\int \textrm{d}{\bf c} \, c^n \, J[f,f]\,\,.
\end{equation}
Recalling the definition of the cumulants (\ref{cumulants}), and the
approximated solution for the stationary velocity pdf, it appears natural to
argue that, for $\lambda \sim 0$, the function $f(\bc, \lambda)$ should be
well approximated by an expansion around the Gaussian, in Sonine polynomials:
\begin{equation}
f(\bc, \lambda)=\phi(c)\left(1+a_1(\lambda) S_1 \left(c^2 \right)
+ a_2(\lambda) S_2 \left(c^2 \right) \right) + {\cal O}(a_3)\, \,,
\end{equation}
where $\phi(c)=\pi^{-d/2} \exp (-c^2)$ is the Gaussian distribution.  Even in
this case, from the relation (\ref{coefficients}) and from the definition
(\ref{v02lambda}), the coefficient $a_1$ is found to be 0.  The method
consists in taking the equation (\ref{moments}) for $n=0$, 2 and 4 in order to
find an explicit expression of $\mu$, $v_0$, and $a_2$ in the limit $\lambda
\to 0$.  The quantities $\nu_2$ and $\nu_4$ have been calculated at the first
order in $a_2$ \cite{vannoijeernst}, and their explicit expressions are:
\begin{equation}
\nu_2={(1-\alpha^2) \over 2 \ell} {\Omega_d \over \sqrt{2 \pi}} 
\left\{ 1 + {3 \over 16} a_2 \right\}= {d \Gamma \over \sqrt{2 T_0^3}} 
\left\{ 1 + {3    \over 16} a_2 \right\} \,, 
\end{equation}
and
\begin{equation}
 \nu_4= {d \Gamma \over \sqrt{2 T_0^3}} \left\{ T_1 + a_2 T_2 \right \},
\end{equation}
with
\begin{align}
  T_1 &=d+{3 \over 2} + \alpha^2 \\
  T_2 &={3 \over 32} (10 \, d + 39 + 10 \, \alpha^2) +{ (d-1) \over (1-
    \alpha)}\,,
\end{align}
where $T_0=\bigg({2 d \Gamma \ell \sqrt{\pi} \over (1-\alpha^2)
  \Omega_d}\bigg)^{2/3}$ is the the granular temperature obtained averaging
over Gaussian velocity pdfs (i.e. the zero-th order of Sonine expansion). The
expression of the first moments $m_n$ is:
\begin{subequations}
\begin{gather}
m_0=1 \\ m_2=d/2 \\ m_4=\frac{\left( 1 + a_2 \right) \,d\,
    \left( 2 + d \right) }{4} \\
m_6=\frac{ \left( 1 + 3\, a_2 \right) \,d\,
      \left( 2 + d \right) \,\left( 4 + d \right)  }
    {8}
\end{gather}
\end{subequations}
With the help of the above defined temperature scale $T_0$, we shall now
introduce some dimensionless variables:
\begin{equation}
\begin{split}
  {\tilde \mu} = \mu {T_0 \over d \Gamma}\,, \quad \quad \quad & \quad \quad
  \quad {\tilde
    \lambda}  = \lambda T_0 \,,\\
  {\tilde v}_0^2 = {v_0^2 \over 2 T_0}\,,\quad \quad \quad & \quad \quad \quad
  {\tilde \nu}_p   = {\sqrt{2 \, T_0^3} \over \Gamma} \nu_p\,.\\
\end{split}
\end{equation}
Note that this scaling naturally defines the scales for the other quantities
of interest, namely: 
\begin{equation}
\begin{split}
  \pit_t=\pi_t {T_0 \over d \Gamma}\,,\quad\quad\,\,\wt={w \over
    d\Gamma}\,,\quad\quad\,\,\cWt={\cW \over \langle \cW \rangle}\,\,.
\end{split}
\end{equation}
The expression of the moment equation (\ref{moments}) becomes, for the above
defined dimensionless quantities:
\begin{equation}
\label{recursiverescaled}
\left( \mut d +(2n+d) \lambt \right) m_n = {n(n+d-2) \over 2 \vt^2}m_{n-2}+
2 \lambt^2 \vt^2  m_{n+2} - \vt {\tilde \nu}_n \,\,.
\end{equation}
First we solve the above equation for $n=0$, getting the following result:
\begin{equation}
\label{solsys1}
{\tilde \mu ({\tilde \lambda})} =- {\tilde \lambda} + {\tilde \lambda}^2 
{\tilde v}_0^2({\tilde \lambda}).
\end{equation}
Recalling that when $\lambda \to 0$ one has $v_0^2 = 2 T_g + {\cal
  O}(\lambda)$, it is important to note that if we restrict our
analysis to the Gaussian approximation for $P({\cal W},t)$, that is if we
truncate $\mu (\lambda)$ to order $\lambda^2$, Eq. (\ref{solsys1}) will
read:
\begin{equation}
\label{secondorder}
{\mu \over d \Gamma} = \lambda (\lambda T_g -1)\,\,.
\end{equation}
Then we see that indeed
\begin{equation}
\label{symGC}
\mu(\lambda)=\mu\left({1 \over T_g}-\lambda\right)\,\,,
\end{equation}
which means that $\pi_{\infty}(w)=\text{max}_{\lambda} \{\mu(\lambda)+\lambda
w\}$ verifies
\begin{equation} \label{pi_secondorder}
\pi_{\infty}(w)-\pi_{\infty}(-w)={w  \over T_g}\,\,.
\end{equation}
This is, up to a prefactor $T_g$, the GCFR.
However, the nontrivial functions $m_n(\lambda)$ will break the
property (\ref{symGC}), as we shall explicitly show later.  In order
to characterize more precisely the dependence of $\mut$ upon $\lambt$
for small values of $\lambt$ , it is useful to expand ${\tilde v}_0^2$
and $a_2$ in powers of ${\tilde \lambda}$:
\begin{subequations}
\label{lambdaexpansion}
\begin{gather}
  {\tilde v}_0^2({\tilde \lambda}) ={\tilde v}_0^{2^{(0)}}+{\tilde
    \lambda} {\tilde v}_0^{2^{(1)}}+{\tilde \lambda}^2 {\tilde v}_0^{2^{(2)}}
 + \cO (\lambt^3)\\
  a_2(\tilde{\lambda}) =a_2^{(0)}+{\tilde \lambda} a_2^{(1)}+{\tilde
    \lambda}^2 a_2^{(2)}+ \cO (\lambt^3)
\end{gather}
\end{subequations}
In this way we can find ${\tilde v}_0^{2^{(i)}} \left(a_2^{(i)}
\right)$ solving equation (\ref{recursiverescaled}) for $n=2$:
\begin{equation}
{\tilde v}_0^{2^{(0)}}= \left(1- {a_2^{(0)}\over 8} \right)\,\,,
\end{equation}

\begin{equation}
{\tilde v}_0^{2^{(1)}}=-{4 \over 3} + {a_2^{(0)} \over 3} -
{a_2^{(1)} \over 8}\,\,,
\end{equation}

\begin{equation}
{\tilde v}_0^{2^{(2)}}= 2 - a_2^{(0)} \left({1 \over 12} + {d \over 3} \right)
+ {a_2^{(1)} \over 3}- {a_2^{(2)} \over 8} \,\,.
\end{equation}
Then we substitute ${\tilde v}_0^2 ({\tilde \lambda})$ in the third equation
and expand it in powers of ${\tilde \lambda}$ to find the expression of
$a_2^{(i)}(\alpha)$. Note that one has also to expand in powers of $a_2$ and
keep only the linear terms in order to be coherent with the ${\tilde \nu}_p$-s
calculations.  We find the following expressions, which are plotted in Fig.
\ref{a2plot}:

\begin{equation} \label{a20}
a_2^{(0)}=\frac{4\,\left( 1 - \Mvariable{\alpha} \right) \,
    \left( 1 - 2\,{\Mvariable{\alpha}}^2 \right) }
  {19 + 14\,d - 3\,\Mvariable{\alpha}\,
    ( 9 + 2d ) + 6 ( 1 - \Mvariable{\alpha} ) \,
       \alpha^2  }
\end{equation}

\begin{equation} \label{a21}
a_2^{(1)}= -\frac{4\,{\left( 1 - \Mvariable{\alpha} \right) }^2\,
    \left( -1 + 2\,{\Mvariable{\alpha}}^2 \right) \,
    \left( 31 + 2\,{\Mvariable{\alpha}}^2 + 16\,d \right) }
    {(19 + 14\,d - 3\,\Mvariable{\alpha}\,
    ( 9 + 2d ) + 6 ( 1 - \Mvariable{\alpha} ) \,
       \alpha^2  )^2}
\end{equation}

\begin{equation}
a_2^{(2)}={A(\alpha) \over B(\alpha)}
\end{equation}
with

\begin{equation}
\begin{split}
  A(\alpha)& =16\,{\left( -1 + \Mvariable{\alpha} \right) }^2\,\left( -1 +
    2\,{\Mvariable{\alpha}}^2 \right) \, \{ 906 + \Mvariable{\alpha}\,\left[
    -984 + \Mvariable{\alpha}\,\left( 85 + 3\,\Mvariable{\alpha}\, \left( -19
        + 6\,\left( -1 + \Mvariable{\alpha} \right) \,\Mvariable{\alpha}
      \right) \right) \right] + 985\,d + \\& + \Mvariable{\alpha}\,\left[ -951
    + \Mvariable{\alpha}\, \left( -25 + 3\,\Mvariable{\alpha}\,\left( 7 +
        6\,\left( -1 + \Mvariable{\alpha} \right) \,\Mvariable{\alpha} \right)
    \right) \right] \,d + \left( 269 + 3\,\Mvariable{\alpha}\,\left( -75 +
      2\,\Mvariable{\alpha}\,\left( -7 + 3\,\Mvariable{\alpha} \right) \right)
  \right) \,d^2 \} \, ,
\end{split}
\end{equation}
and
\begin{equation}
B(\alpha)=3\,
    {\left( -19 - 14\,d + 3\,\Mvariable{\alpha}\,\left( 9 + 2\,\left( -1 
+ \Mvariable{\alpha} \right) \,\Mvariable{\alpha} + 2\,d \right)  \right) }^3
\end{equation}
The $v_0^{2^{(0)}}$ expression, as well as the $a_2^{(0)}$ expression,
coincide with
the usual results established for granular gases
\cite{vannoijeernst,montanerosantos}.  At this point the computation of the
cumulants becomes straightforward. From relation (\ref{cumulants}) it follows:
\begin{equation}
\lim_{t \to \infty} \frac{\langle \cW^n \rangle_c}{t} = (-1)^n d \Gamma
T_0^{n-1} n! \, \, \vt^{2^{(n-2)}}\,\,.
\end{equation}
Moreover, since the $a_2^{(i)}$ corrections are numerically small, as shown
in Fig. \ref{a2plot}, the zero-th
order (Gaussian) approximation already gives a good estimate for the
cumulants.  Namely, the first cumulants are , in this approximation:
\begin{equation} \label{4cumulants}
\begin{split}
  \left\langle \cW \right\rangle_c = t N d \Gamma\,\,,\quad \quad \quad & \quad
  \quad \quad
  \left\langle \cW^2 \right\rangle_c= 2 t N d \Gamma T_0\,\,,\\
  \left\langle \cW^3 \right\rangle_c = 8 t N d \Gamma T_0^2\,\,,\quad \quad
  \quad & \quad \quad \quad
  \left\langle \cW^4 \right\rangle_c   =  48  t N d \Gamma T_0^3\,\,.\\
\end{split}
\end{equation}
All the above expansions in powers of $\lambda$, at the second order in Sonine
coefficients (e.g. $a_2$) can be carried out by expanding $v_0$ and $a_2$ in
(\ref{lambdaexpansion}) to higher powers of $\lambda$. Moreover, expanding in
higher order in Sonine coefficient (e.g. $a_3$) remains in principle still
possible, but will involve a higher number of equations in the hierarchy
(\ref{recursiverescaled}) (e.g. $n=6$), and therefore will need the expression
of higher order collisional moments (e.g. $\nu_6$).

\subsection{The solvable infinite dimension limit}

The previous discussion strongly suggests that at high dimensions
$\ft(\bv,\lambda)$ is not far from a Gaussian. First, the inspection of the
$d\to\infty$ limit of the Sonine expansion performed in the previous
subsection indicates that the corrections to the Gaussian (see
formula~\ref{a20} and~\ref{a21}) carry a relative $d^{-1}$ factor, thus
leaving the Gaussian pdf as the leading behavior when $d\to\infty$. Second,
the $\lambda=0$ case is well known to reproduce a Gaussian in the $d\to\infty$
limit: there is numerical evidence that the collision integral is solved (i.e.
is set to zero) by a Gaussian $\ft$, when $d\to\infty$. Of course if the
Gaussian solves the collision integral then it solves the whole equation.

We are therefore led to consider, in the limit $d \to \infty$, $\ft(\bv,
\lambda)$ to be a Gaussian with a $\lambda$-dependent second moment. In this
situation the dimensionless function $f$ will read:
\begin{equation}
f(\bc)={e^{-c^2} \over \pi^{d/2}}
\end{equation}
with ${\bf c}(\lambda)={\bf v}/v_0(\lambda)$.  In this context one can solve
equation (\ref{recursiverescaled}) in order to get an explicit expression for
$\mu(\lambda)$.  The equation defined by (\ref{recursiverescaled}) for
$n=0$ gives:
\begin{equation}
\mut (\lambt)=-\lambt + \lambt^2  {\tilde v}_0^2(\lambt)\,,
\label{eq:mudinfini}
\end{equation}
where ${\tilde v}_0^2(\lambt)$ is obtained from Eq. (\ref{recursiverescaled}) for
$n=2$, which reads:
\begin{equation}
\lambt^2 \vt^4 -\vt^3-2 \lambt \vt^2+1=0\,\,\,\,.
\end{equation}
The unique solution of the above equation which verifies the
physical requirement $\vt^2(0)=1$ is:
\begin{equation}
{\tilde v}_0^2(\lambt)= 
\frac{1 + 4 \,{\lambt}^3}{4\,{\lambt}^4} +
{b_1(\lambt) \over 2}-{1 \over 2}
\left [ -\frac{8}{{\lambt}^2} + \frac{{\left( 1 + 4\,{\lambt}^3 
 \right) }^2}{2\,{\lambt}^8}+b_2(\lambt)-b_3(\lambt)+ 
{b_4(\lambt) \over 4 b_1(\lambt)} \right]^{1 \over 2} \,\,,
\end{equation}
with
\begin{equation}
\begin{split}
  b_1(\lambt)= \sqrt{\frac{{\lambt}^{-8}}{4} + \frac{2}{{\lambt}^5} -
    b_2(\lambt)+b_3(\lambt)}\,, \quad & 
    b_2(\lambt)=\frac{4\,{\left(
        \frac{2}{3} \right) }^{\frac{1}{3}}} {{\lambt}^3 \,{\left( 9 +
        {\sqrt{3}}\,{\sqrt{27 +
            256\,{\lambt}^3}} \right) }^{\frac{1}{3}}} \,,\\ 
  b_3(\lambt)=\frac{{\left( 9 + {\sqrt{3}}\,{\sqrt{27 + 256\,{\lambt}^3}}
      \right)}^{\frac{1}{3}}}{2^{\frac{1}{3}}\,3^{\frac{2}{3}}\,{\lambt}^4}\,,
  \quad &  
  b_4(\lambt)=\frac{32}{{\lambt}^3} - \frac{24\,\left( 1 + 4\,{\lambt}^3
    \right) }{{\lambt}^6} +
  \frac{{\left( 1 + 4\,{\lambt}^3 \right) }^3}{{\lambt}^{12}}\,.\\
\end{split}
\end{equation}

\begin{figure}
  \centering
\includegraphics[width=0.5 \textwidth]{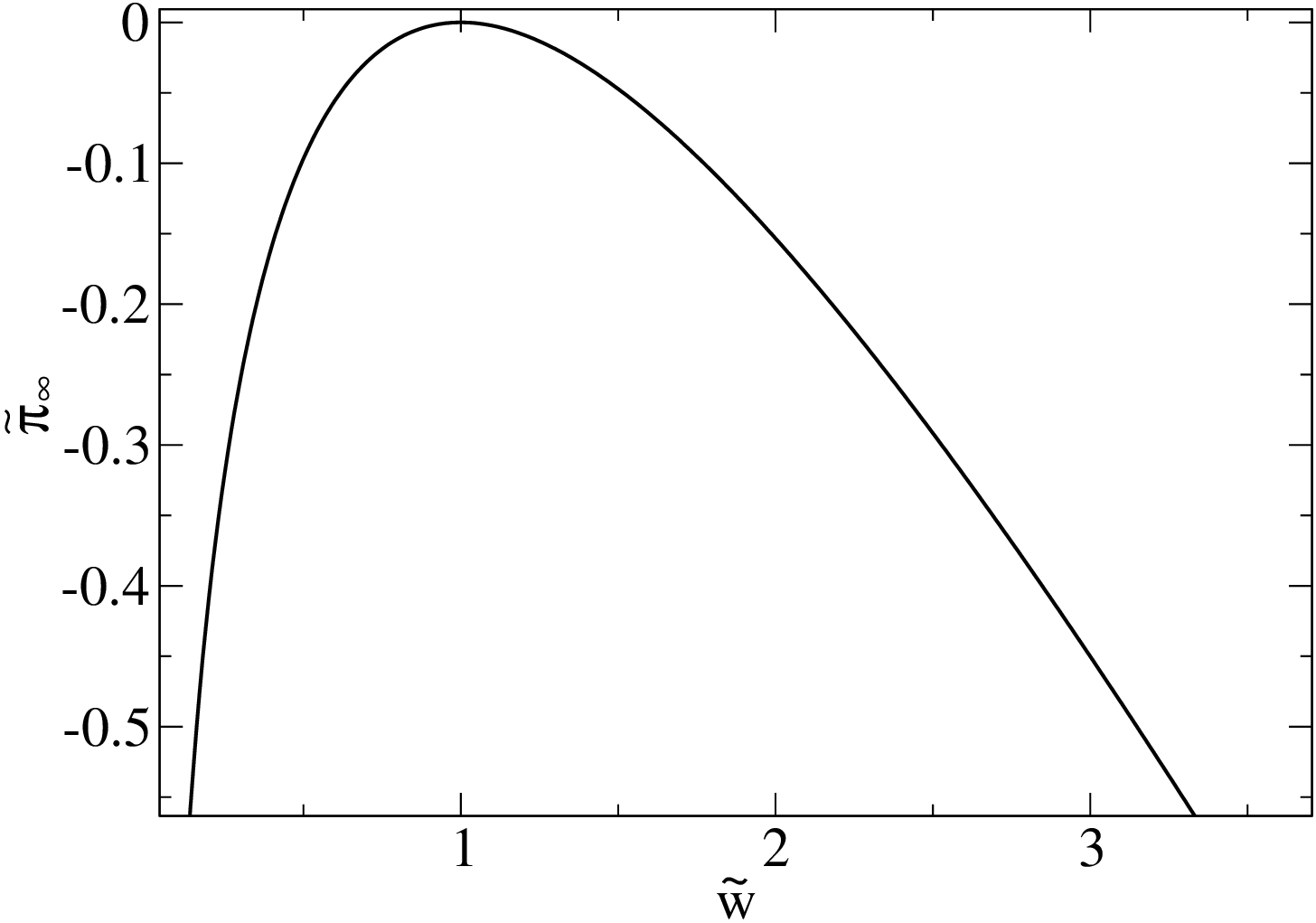} 
\caption{Large deviation function $\pit_{\infty}(\wt)$ for $d=\infty$,
as obtained 
by numerical inverse Legendre transform of Eq. (\ref{eq:mudinfini}).}
\label{piofw} 
\end{figure}

This expression of the velocity scale reduces to the kinetic temperature for
$\lambda=0$, and decreases monotonically as $\lambda^{-1/2}$ when $\lambda \to
\infty$. This means that in the limit $\lambda \to \infty$ $\ft$ approaches a
Dirac distribution as $\exp(-\lambda v^2/2)$. This feature supports the
intuition that the small ${\cal W}$ events (which are related to the large
values of $\lambda$) are provided by the small velocities. The behavior of
$\mut$ is shown in Fig.  \ref{plotmunoappr}.  The large deviations function
$\mut (\lambdat)$ becomes complex for $\lambdat < -{3 \over 2^{8/3}}$, because
of the terms containing $\sqrt{27 + 256 \lambdat^3}$.  Moreover for large
$\lambdat$ the behavior of this function is $\mut (\lambdat) \sim -
\lambdat^{1 \over 4}$.  In the vicinity of the singularity (i.e.  $\lambdat=
\lambda_0 =-{ 3 \over 2^{8/3}}$) the behavior of the large deviation function
is:
\begin{equation}
\mut (\lambdat) = {3\over 2^{2/3}} - 3^{3/2} 2^{1/6} \sqrt{\lambdat -
    \lambda_0} + {\cal O}(\lambdat- \lambda_0)\,\,.
\end{equation}
From the behavior for large $\lambt$ it is possible to recover the left tail
of the large deviation function $\pi_{\infty}$. In general, if $\mu(\lambda)
\sim -\lambda^{\beta}$ for $\lambda \to \infty$, this leads to
$\mu'(\lambda_*)=-\beta \lambda_*^{\beta-1}=-w$. This last relation tells us
that for $\beta<1$ we are recovering the limit $w\to 0^+$, with a behavior of
the large deviation function given by $\pi_{\infty}(w)=\mu(\lambda_*)+
\lambda_* w\sim w^{\beta \over \beta-1}$.  Moreover, from the behavior of
$\mu$ near $\lambda_0$, an analogous calculation provides the right tail of
the large deviation function: $\pi_{\infty}(w) \sim \lambda_0 w$, when $w \to
\infty$.  Finally, in our particular case, the tails are given by
\begin{equation}
\pit_{\infty}(\wt \to 0^+) \sim -\wt^{-1/3}\,\,,
\,\,\pit_{\infty}(\wt\to \infty) \sim -\wt\,\,,
\end{equation}
Note that, as expected from the discussion in subsection \ref{EnergyBalance},
there is no $w<0$ tail to $\pit_{\infty}$. The graph of the whole function
$\pit_{\infty}(\wt)$ is depicted in Fig. \ref{piofw}.

\section{Inelastic Maxwell gas}
\label{sec:Maxwell}

\subsection{The model}

The Maxwell gas is a kinetic model due to J.C. Maxwell, who observed that a
pair potential proportional to $r^{-2(d-1)}$, being $r$ the distance between
two interacting particles, gives rise to a great simplification of the
collision integral \cite{maxwell}. In fact this kind of interaction makes the
collision frequency velocity independent. It must be noted that when the
inelasticity of the particles is considered, this model looses its
well-defined physical interpretation, but it nevertheless keeps its own
interest. The collision integral is analytically simpler than the hard
particles model and preserves the essential physical ingredients in order to
have qualitatively the same phenomenology. In the recent development of
granular gases this kinetic model has been extensively investigated
\cite{puglisi,bennaimkrap,britoernst,bobylev}. In particular Inelastic Maxwell
molecules display strong non-Gaussian velocity pdfs (generally stronger than
for the hard spheres), which can be even pathological in some cases. For
example the velocity pdf of the cooling regime of such a gas has tails with
algebraic decay.  Nevertheless for the stochastically driven gas such a
pathology does not exists, and the tails of the velocity pdf have an
exponential decay. Moreover, as it will appear in the next section, with this
model the DSMC simulation algorithm is highly optimized, and thus a much more
precise observation of rare events can be achieved.

The Boltzmann equation governing the single particle velocity distribution is:
\begin{equation}
\label{boltzmax}
\partial_t f(\bv_1, t) = J_M[f,f] +\Gamma \Delta_{\bv_1} f(\bv_1)\,\,,
\end{equation}   
where $J_M$ is the collision integral, which slightly changes from
(\ref{collintboltz}):
\begin{equation}
J_M[f,f]=\omega_0 \int \dd \bv_2 \int \dd \bsig 
\left( {f(\bv_1^{**}) f(\bv_2^{**})\over \alpha} - 
f(\bv_1) f(\bv_2) \right)\,.
\end{equation}
The solution of such equation have been extensively analyzed
\cite{puglisi,bennaimkrap,britoernst}, and it turns out that
even in this case a Sonine expansion up to the second order provides a good
estimate of the pdf for low velocities, while the tails of the pdf are
exponential.

In order to compute the large deviation function associated to the total work
distribution, the same reasoning carried out in the previous part still
holds. Thus, the analogous of equation (\ref{lambdaBoltzmann}) becomes, for
Maxwell molecules:
\begin{equation}\label{lambdaMaxwell}\begin{split}
    \mu\tilde{f}(\bv,\lambda,t)=&\Gamma\Delta_{\bv}\tilde{f}+2\lambda \Gamma
    \bv\cdot\p_{\bv}\tilde{f}+\Gamma(d\lambda+\lambda^2 v^2)\tilde{f}\\&+
    \omega_0 \int \dd \bv_2 \int \dd\bsig
    \left[\alpha^{-1}\tilde{f}(\bv_1^{**},\lambda)\tilde{f}(\bv_2^{**},\lambda)
      -\tilde{f}(\bv_1,\lambda)\tilde{f}(\bv_2,\lambda)\right]\,\,,
\end{split}
\end{equation}
where the same molecular-chaos-like assumption as in 
section \ref{sec:IHS} has been implemented.  

\subsection{The cumulants}
\begin{figure}
\begin{minipage}[t]{.46\linewidth}
 \includegraphics[width=1 \textwidth]{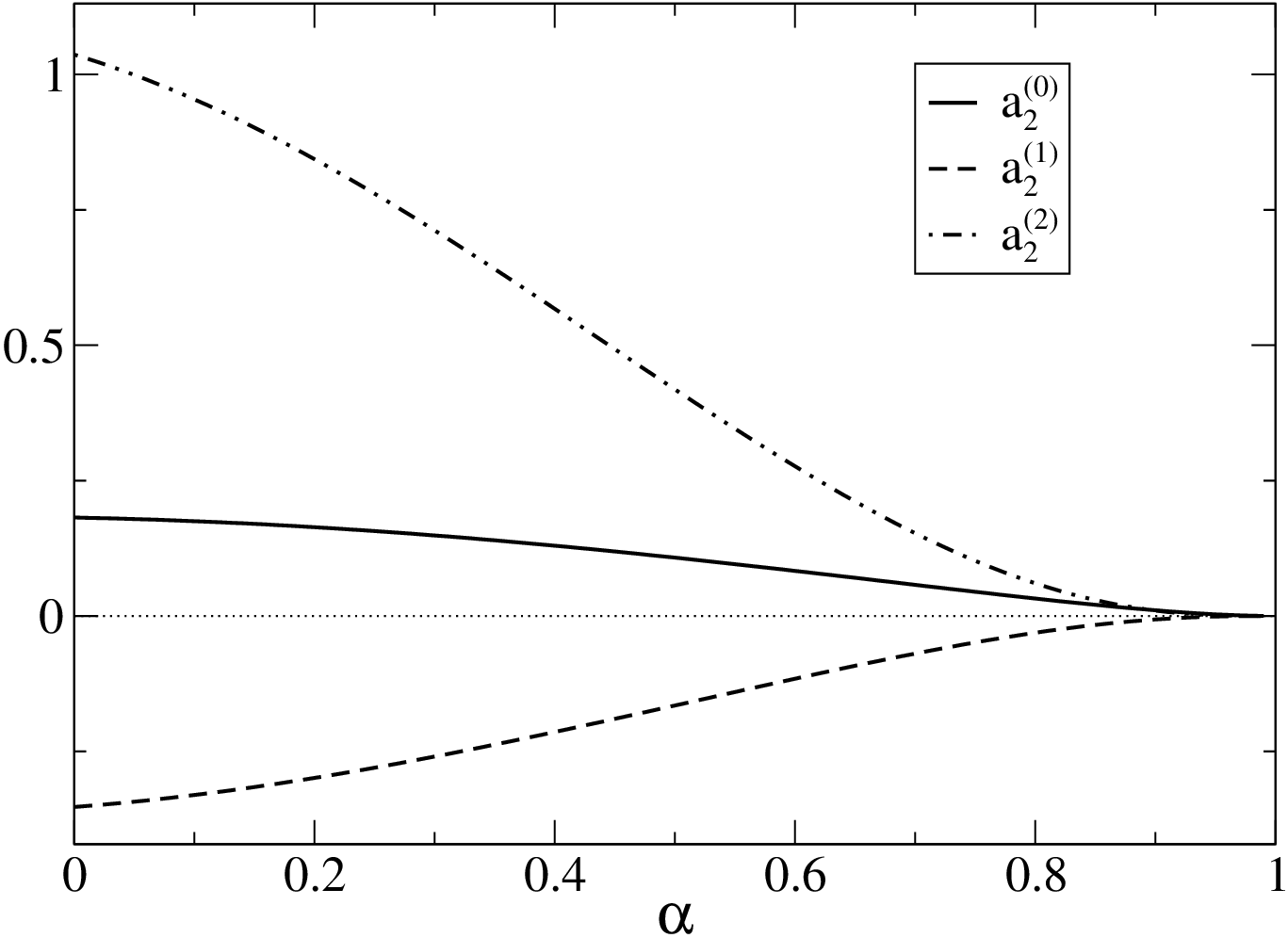}
\caption{\label{a2plotmax}
$a_2^{(0)}$, $a_2^{(1)}$ and $a_2^{(2)}$ versus $\alpha$ for the 
inelastic Maxwell model
in $d=2$.}
\end{minipage}
\hfill
\begin{minipage}[t]{.46\linewidth}
  \includegraphics[width=1 \textwidth]{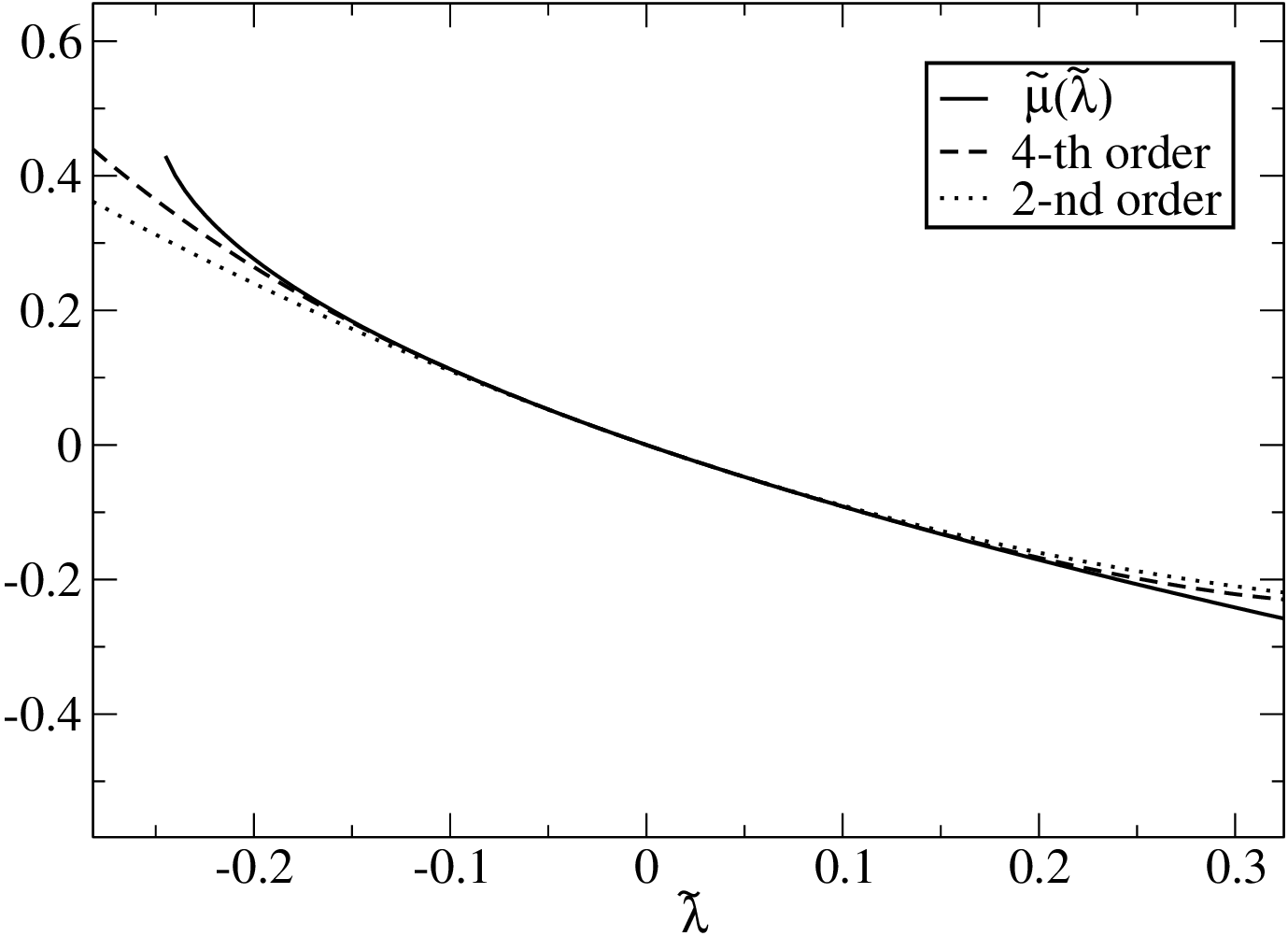}
\caption{\label{plotmunoapprmax}The solid line shows $\mut$ in the
  limit $d \to \infty$ for the inelastic Maxwell model, given in Eq.
  (\ref{muMaxinf}). The dashed line is $\mut$ at fourth order in $\lambt$
  from (\ref{muMax}) for $d=2$ and $\alpha=0.5$. Finally the dotted line
  shows the same quantity calculated with a truncation at second order in
  $\lambda$, which would satisfy the GCFR. }
\end{minipage}
\end{figure}

In this section we will provide the computation of the cumulants for the
integrated power injected by the thermostat. The procedure is formally the
same as in the previous section, for the hard sphere gas. The recursive
equation involving the velocity moments is found from (\ref{lambdaMaxwell}),
and reads:
\begin{equation}
\label{maxmom}
(\mu + \Gamma(2n+d) \lambda)m_n={\Gamma \over v_0^2} n(n+d-2)m_{n-2}
+\Gamma \lambda^2 v_0^2 m_{n+2}- \omega_0  \nu_n\,\,,
\end{equation}
where $m_n$ is the $n-$th moment related to the dimensionless function $f$
(cf. eq (\ref{remoments})).
The expression of the collisional moments is now slightly different. The two
first non-vanishing moments are \cite{santos}:
\begin{equation}
\nu_2= {1- \alpha^2 \over 4} = {d \Gamma \over \omega_0 T_0}\,\,,
\end{equation}
and
\begin{equation}
\nu_4= {\nu_2 \over 8} (T_1 + a_2 T_2)\,\,,
\end{equation}
with
\begin{equation}
T_1 = 2 (5+3 \alpha^2 +4 d)\,\,,
\end{equation}
\begin{equation}
T_2 = 3 \alpha^2 -{\alpha (17+4 d) -3 (3+4 d) \over 1-\alpha}\,\,.
\end{equation}
Here again $T_0= {4 d \Gamma \over \omega_0 (1-\alpha^2)}$ denotes the zero-th
order temperature. Let us introduce the following rescaled variables:
\begin{equation}
\label{rescaling_maxwell}
\begin{split}
  {\tilde \mu} = \mu {T_0 \over d \Gamma} \quad \quad \quad & \quad \quad
  \quad {\tilde
    \lambda}  = \lambda T_0 \\
  {\tilde v}_0^2 = {v_0^2 \over 2 T_0}\quad \quad \quad & \quad \quad \quad
  {\tilde \nu}_p   = {\omega_0 \, T_0 \over d \Gamma} \nu_p\\
\end{split}
\end{equation}
The equation for the moment reads for the dimensionless variables:
\begin{equation}
\label{moments_maxwell}
(d {\tilde \mu} + (2n+d){\tilde \lambda})m_n=
{1 \over 2 {\tilde v}_0^2} n (n+d-2) m_{n-2} + 
2 {\tilde \lambda}^2 {\tilde v}_0^2 m_{n+2} -d {\tilde \nu}_n\,\,.
\end{equation}
Expanding again the solution in a Gaussian distribution multiplied by a series
of Sonine polynomials up to the second order one obtains, from the first three
equations of the hierarchy (\ref{moments_maxwell}), a closed set of equations
involving $\mut$, $\vt^2$ and $a_2$. Those equations can be solved in the
vicinity of $\lambda=0$ and in the limit of small non-Gaussianity
($|a_2(\lambda)|\ll 1$), when linearized with respect to the coefficient $a_2$.
In this framework, which should be valid in the limit of small values of
$\lambt$, the solution is:
\begin{equation}
\label{muMax}
\mu(\lambda)= \frac{1 - {\sqrt{1 + 4\,\lambda}}}{2} - 
  a_2 \, \frac{\left( 2 + d \right) \,
     \left( -1 + {\sqrt{1 + 4\,\lambda}} + 
       2\,\lambda\,\left( -2 - \lambda + 
          {\sqrt{1 + 4\,\lambda}} \right)  \right) }
     {4\,{\sqrt{1 + 4\,\lambda}}} + {\cal O}(a_2^2) \,\,
\end{equation}
with
\begin{equation}
a_2(\lambda)=\frac{-6\,{\left( 1 - \alpha \right) }^2\,
    \left( 1 + \alpha \right) }{\left( 1 + \alpha
       \right) \,\left( 7 + 
       3\,\left(\alpha -2 \right) \,\alpha - 4\,d
       \right)  + 8\,\left( \alpha -1 \right) \,
     \left( 2 + d \right) \,{\sqrt{1 + 4\,\lambda}}}\,\,.
\end{equation}
Expanding this expression in powers of $\lambt$, it is possible 
to recover all the
coefficients $a_2^{(i)}$ of the expansion (\ref{lambdaexpansion}). The first
coefficients are plotted in Fig. \ref{a2plotmax}. 
These results allows us to directly compute the cumulants from the
derivatives of $\mut$:
\begin{equation}
\langle {\cal W}^n \rangle_c = d \Gamma T_0^{n-1} \left.{\dd^n \mut \over \dd
  \lambt^n}\right|_{\lambt=0} \,\,.
\end{equation}
It turns out that the corrections coming from the Sonine expansion
are small with respect to the Gaussian order, which therefore
already provides a good estimate. The values of the first cumulants are:
\begin{equation}
\begin{split}
  \left\langle \cW \right\rangle_c = t N d \Gamma\,\,,\quad \quad \quad &
  \quad \quad \quad
  \left\langle \cW^2 \right\rangle_c= 2 t N d \Gamma T_0\,\,,\\
  \left\langle \cW^3 \right\rangle_c = 12 t N d \Gamma T_0^2\,\,,\quad \quad
  \quad & \quad \quad \quad
  \left\langle \cW^4 \right\rangle_c   =  120  t N d \Gamma T_0^3\,\,.\\
\end{split}
\end{equation}

\subsection{The solvable infinite dimension limit}
\begin{figure}
  \centering
\includegraphics[width=0.5 \textwidth,clip=true]{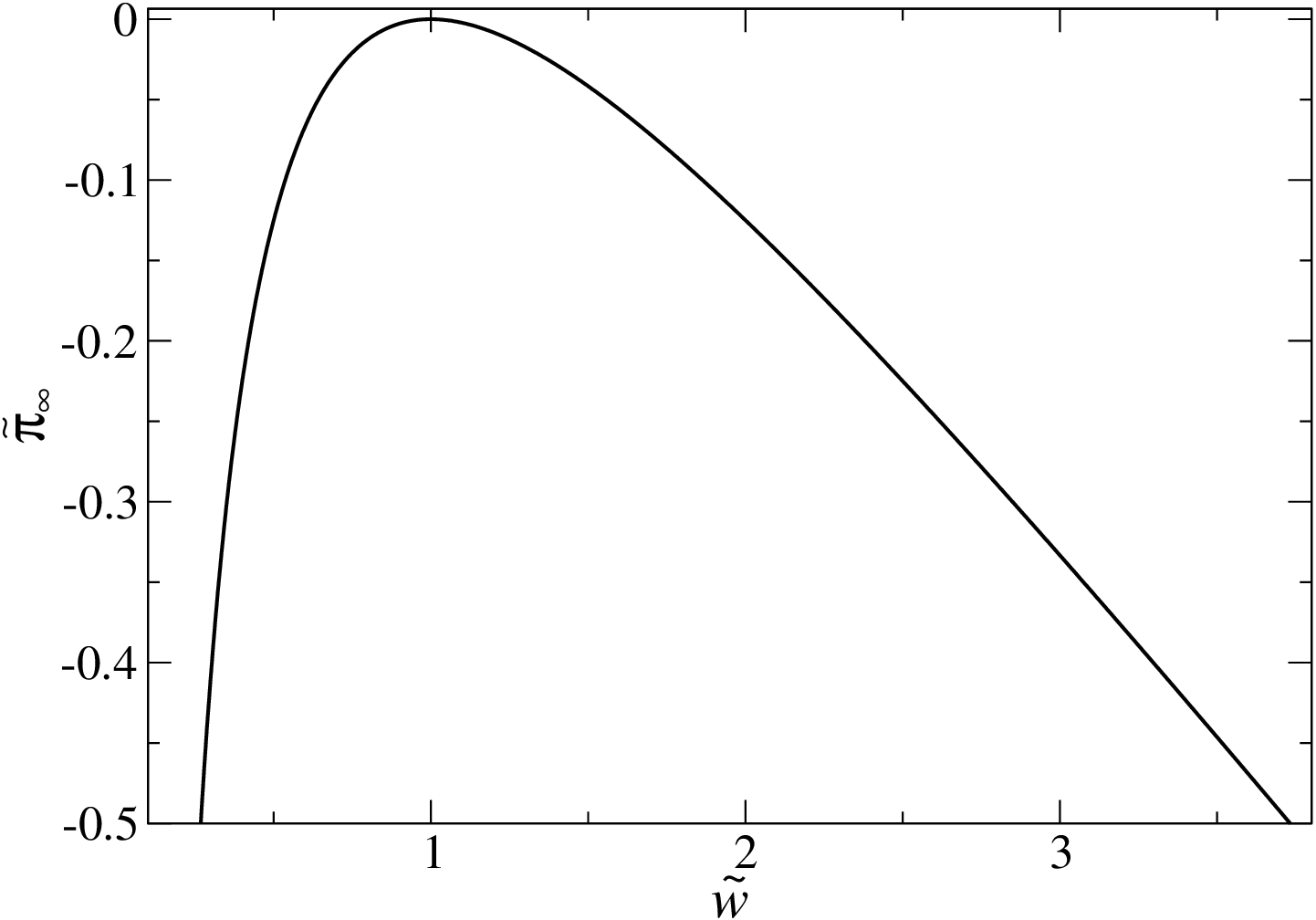} 
\caption{$\pit_{\infty}(\wt)$ for the Inelastic Maxwell Model in the
limit $d \to \infty$.}
\label{piofwmax} 
\end{figure}

In this section we will provide an exact solution for the generating function
$\mu(\lambda)$ in the infinite dimension limit. As already noted by Ben-Naim
and Krapivsky, in this limit the velocity pdf approaches the Gaussian
distribution~\cite{bennaimkrap}. 
Therefore, we will show that a Maxwellian with a variance
depending on $\lambda$ is indeed a solution of equation
(\ref{lambdaMaxwell}). Because of the convolution structure of the collision
integral, a great simplification occurs introducing the Fourier Transform of
the velocity pdf:
\begin{equation}
F(\bk,\lambda)=\int \dd \bv \,\ee^{i \bk \cdot \bv} f(\bv, \lambda) \,.
\end{equation}
Assuming that $F$ is an isotropic function of the wave vector $\bk$, 
the equation (\ref{lambdaMaxwell}) reads:
\begin{equation}
\label{lambdaFourier}
\mu F(k)=- \Gamma k^2 F(k) -\lambda \Gamma d F(k)-2 \lambda \Gamma k 
{\p F(k) \over \p k}- \lambda^2 \Gamma {\p^2 F(k) \over \p k^2} + 
\omega_0 \left\langle F(\sqrt{\xi} k) F(\sqrt{\eta}k) 
\right\rangle_{\theta} - \omega_0 F(k)
\end{equation}
where
\begin{equation}
\xi=1-\left( 1-\left({1-\alpha \over 2} \right)^2 \right) \theta 
\,\,\,\,,\,\,\,\,
\eta=\left({1+\alpha \over 2}\right)^2  \theta \,\,,
\end{equation}
and the brackets with a subscript $\theta$ denote an angular average
\begin{equation}
\langle f \rangle_{\theta} = \int_0^1 d\theta {\theta^{-{1 \over 2}} 
(1-\theta)^{d-3 \over 2}
  \over B\left({1\over 2}, {d-1 \over 2} \right)} f(\theta)\,\,,
\end{equation}
where $B$ is the beta function. Injecting a Gaussian solution
$\phi(k)=\ee^{-{k^2 T \over 2}}$ into the above equation yields the following
relation:
\begin{equation}
\mu = - \Gamma k^2 - \lambda d \Gamma +2 \lambda k^2 T + \lambda^2 \Gamma T -
\lambda^2 \Gamma k^2 T^2 + \omega_0 \left\langle \ee^{-{k^2 (1-\alpha^2)
      \theta T  \over 4}} \right\rangle_{\theta} - \omega_0\,.
\end{equation}
Then, expanding the exponential term in a power series and using that
\begin{equation}
\langle \theta^p \rangle_{\theta} ={\Gamma({d \over 2}) \Gamma(p+{1\over 2}) 
\over \Gamma({1 \over 2}) \Gamma(p+{d \over 2})} \sim {1 \over d^p}\,,
\end{equation} 
when $d \to \infty$, one sees that the above relation holds for each $k$,
with:
\begin{equation}
\mu(\lambda)=-\lambda d \Gamma +\lambda^2 \Gamma T(\lambda)
\end{equation}
and,
\begin{equation}
T(\lambda)= \frac{1 + 2\,\lambda\,\Mvariable{T_0} - 
 {\sqrt{1 + 4\,\lambda\, \Mvariable{T_0}}}}{2\,
 {\lambda}^2\,\Mvariable{T_0}}
\end{equation}
where $T_0$ is the granular temperature.
Hence, the explicit form of the generating function is, using the rescaled
variables defined in (\ref{rescaling_maxwell}) is:
\begin{equation}
\label{muMaxinf}
\mut (\lambt)= \frac{1 - {\sqrt{1 + 4\,\lambt}}}{2} \,\,.
\end{equation}
This function has the same behavior as its analogous for the IHS gas, since it
presents a cut in the negative real axis (for $\lambda<-1/4$), and it
decreases monotonically (as $-\lambda^{1/2}$) when $\lambda \to \infty$.
Furthermore this expression is, up to irrelevant constants, exactly the same
as the one found in a simpler system, the Ornstein-Ulhenbeck process, which
has been extensively analyzed by Farago in \cite{farago,farago2}.  The
analytical expression of the large deviation function for the total work is
easily computed as the Legendre transform of the above expression of $\mut
(\lambt)$. It reads:
\begin{equation}
\pit_\infty(\wt)=-{(\wt-1)^2 \over 4 \wt} \,\Theta(\wt)\,\,.
\end{equation}
This function is plotted in Fig. \ref{piofwmax}. One can easily see that even
for this model there are no negative events in the large time limit.  The left
tail of the large deviation function decrease to $- \infty$ as $-1/4\wt$,
while the right tail has a linear decreasing behavior.

\section{Numerical results}
\label{sec:numerics}

In this section the results of numerical simulations of the two models
(inelastic hard spheres and inelastic Maxwell model) are presented with
particular attention to the Fluctuation Relation for the
injected power. We emphasize that obviously only $\pi_t(w)$ can be accessed
numerically whereas the GCFR bears on $\pi_\infty(w)$. This requires
a precise discussion of finite time effects, that play a crucial 
role here.
The main requirement to verify the validity of the Fluctuation
Relation is a clean observation of a negative tail in the pdf of the injected
power. This poses a dramatic limit to the time $t$ of integration of $\cW(t)$.
In numerical simulations, as well as in real experiments, at times larger than
a few mean free times the negative tail disappears. On the other hand, at
times of the order of $1$-$3$ mean free times, the Fluctuation Relation
appears to be correctly verified for the inelastic Hard Spheres model and
slightly violated for the inelastic Maxwell model. The measure of the
cumulants, anyway, gives a neat indication of the fact that the time of
convergence of the large deviation function is at least $10$ times larger
and that the true asymptotic is well reproduced by the theory exposed in this
article.

The stationary state of a driven granular gas, modeled by
equation~\eqref{boltz}, under the assumption of Molecular Chaos, is very well
reproduced by a Direct Simulation Monte Carlo~\cite{bird,montanerosantos}.  As
a first check of reliability of the algorithm, we have measured the granular
temperature $T_g$ and the first non-zero Sonine coefficient $a_2\equiv
(\langle v^4 \rangle/\langle v^2 \rangle^2-3)/3$.  The measured granular
temperature is always in perfect agreement with the estimate. The measured
$a_2$ coefficient is a highly fluctuating quantity and its average is in very
good agreement with the theoretical estimate.

\subsection{Inelastic Hard spheres}

\begin{figure}
\begin{minipage}[t]{.46\linewidth}
  \includegraphics[width=1 \textwidth,clip=true]{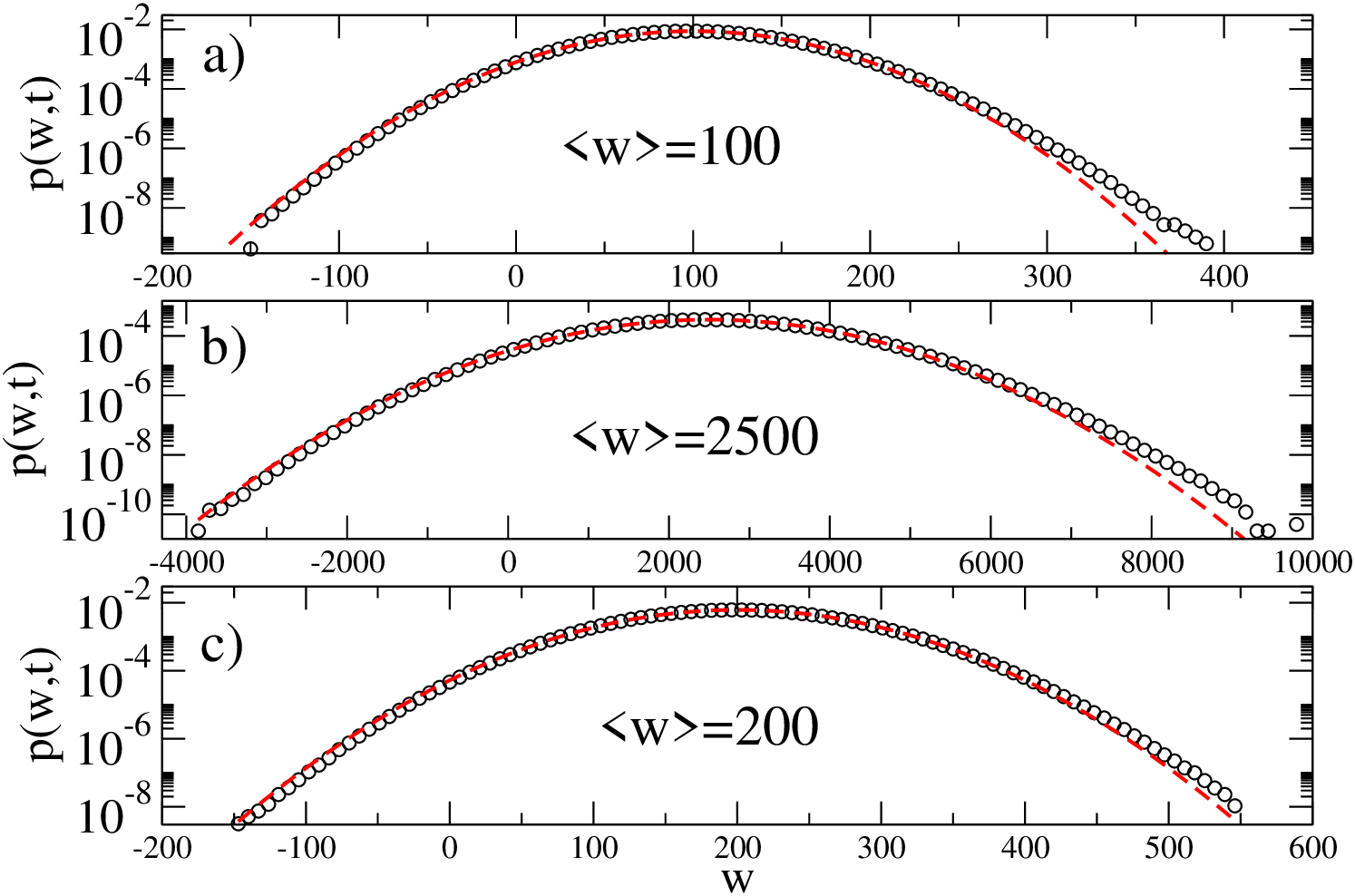}
\caption{Probability density function of the injected power, $p(w,t)\equiv
  tP({\cal W}(t)=wt,t)$ with $t$ equal to $1$ mean free time. In all three
  cases the value of the restitution coefficient is $\alpha=0.9$. Other
  parameters are a) $N=100$, $\Gamma=0.5$; b) N=100, $\Gamma=12.5$; c)
  $N=200$, $\Gamma=0.5$. The dashed line represents a Gaussian with the 
  same first
  two cumulants. These distributions have been obtained with $\sim 1.5 \times
  10^9$ independent values of ${\cal W}(t)$.}\label{fig:pdf}
\end{minipage}
\hfill
\begin{minipage}[t]{.46\linewidth}
  \includegraphics[width=1 \textwidth,clip=true]{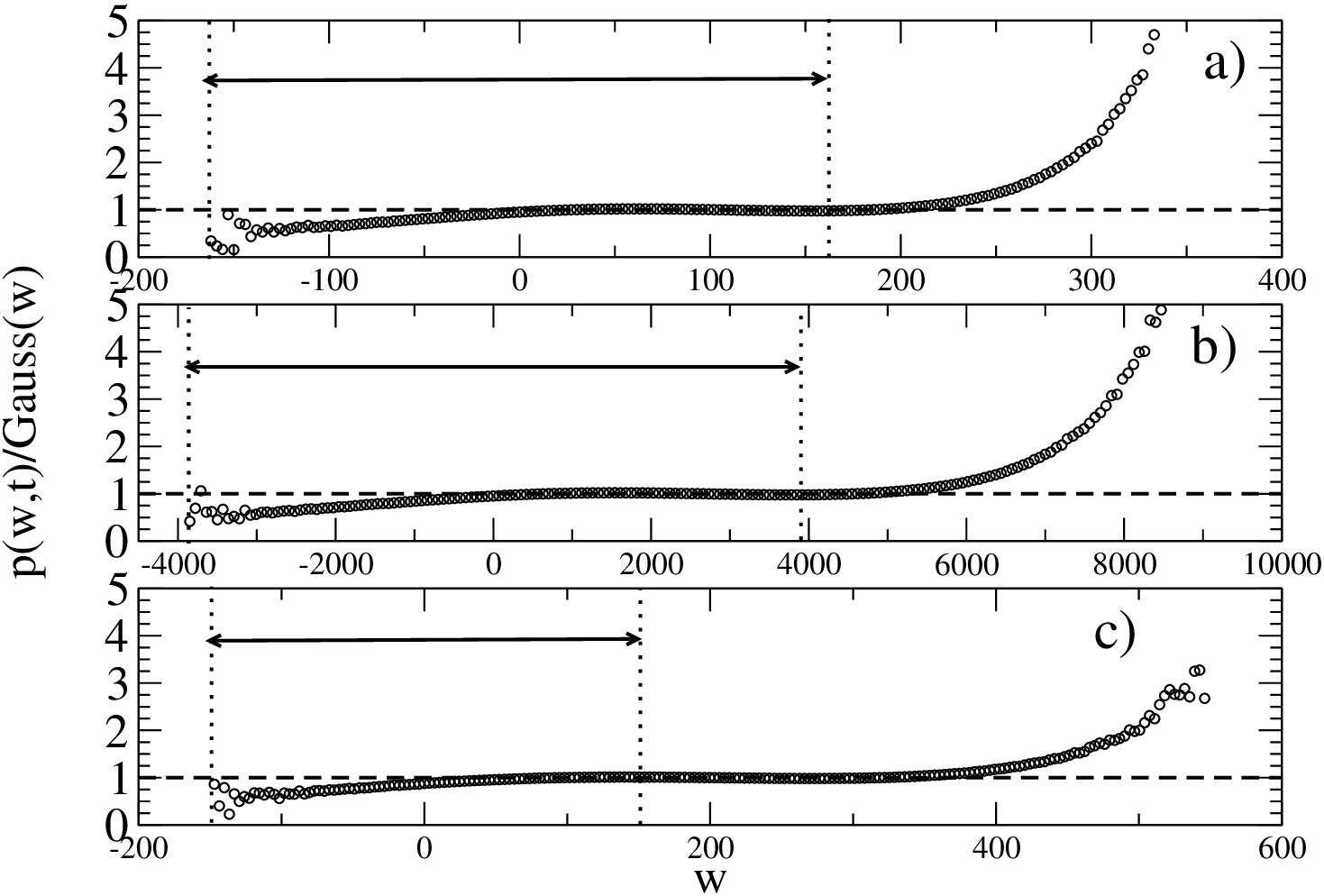}
\caption{Ratio of $P({\cal W},t)$ and a Gaussian with the same first two 
  moments, for the same parameters as in figure~\ref{fig:pdf}: a) corresponds
  to $N=100$, $\Gamma=0.5$, b) to $N=100$, $\Gamma=12.5$ and c) to $N=200$,
  $\Gamma=0.5$. The range between the vertical dotted lines is the useful one
  for the check of the Gallavotti-Cohen relation. It can be noted that the
  strongest deviations from the Gaussian behavior appear outside of this
  range.}\label{fig:pdfratio}
\end{minipage}
\end{figure}

In Fig.~\ref{fig:pdf}, the probability density functions $p(w,t)\equiv
tP(wt,t)$ (only for $t$ equal to $1$ mean free time) for three different
choices of parameters $N,\Gamma$ (at fixed restitution coefficient $\alpha$)
is shown. The values of the first two cumulants of the distribution and their
theoretical values are compared in table~\ref{tab:pdf}, with very good
agreement. In the same table we present also the measure of the third and
fourth rescaled cumulants. The formulae for the first few cumulants are:

\begin{subequations}
\begin{align}
  \langle x \rangle_c &= \langle x \rangle \\
  \langle x^2 \rangle_c &= \langle x^2 \rangle - \langle x \rangle^2 \\
  \langle x^3 \rangle_c &= \langle x^3 \rangle - 3\langle x^2\rangle\langle x
  \rangle+ 2\langle x \rangle^3\\
  \langle x^4 \rangle_c &= \langle x^4 \rangle - 4\langle x^3\rangle\langle x
  \rangle + 12 \langle x^2 \rangle \langle x \rangle^2 - 6 \langle x \rangle^4
  -
  3 \langle x^2 \rangle^2\\
\end{align}
\end{subequations}

\begin{table}[ht]
\begin{tabular}{|l|l|l|l|l|l|l|l|}
  N   &$\Gamma$  &$\langle {\cal W}(t) \rangle/t$ &$\langle {\cal W}(t)^2
  \rangle_c/t$ &$N \Gamma d$
  &$2 N \Gamma d T_g$ & $\sqrt{t} \,\, \sigma(t)$ &$t \,\, \kappa(t) $\\ \hline
  100  &0.5    &100  &20835 &100 &21052 &$0.200432$
  &$0.355176$\\ \hline
  100  &12.5   &2500 &13019125 &2500 &13157900
  &$0.201739$ &$0.361635$\\ \hline
  200  &0.5    &199.9  &42009  &200  &42120
  &$0.141589$ &$0.175454$
\end{tabular}
\caption{First cumulants, 
together with the skewness $\sigma(t)= \langle
  {\cal W}(t)^3 \rangle_c/ \left(\langle {\cal W} (t)^2 \rangle_c \right)^{3/2}$
and the kurtosis excess 
$\kappa(t)=\langle{\cal W}(t)^4 \rangle_c/\left(\langle {\cal W}(t)^2 \rangle_c\right)^{2}$ 
of the distribution of injected work $P({\cal W},t)$, measured with $t$ equal
to $1$ mean free time for different choices of the parameters.}\label{tab:pdf}
\end{table}

\begin{figure}[ht]
\begin{center}
  \includegraphics[height=8cm,clip=true]{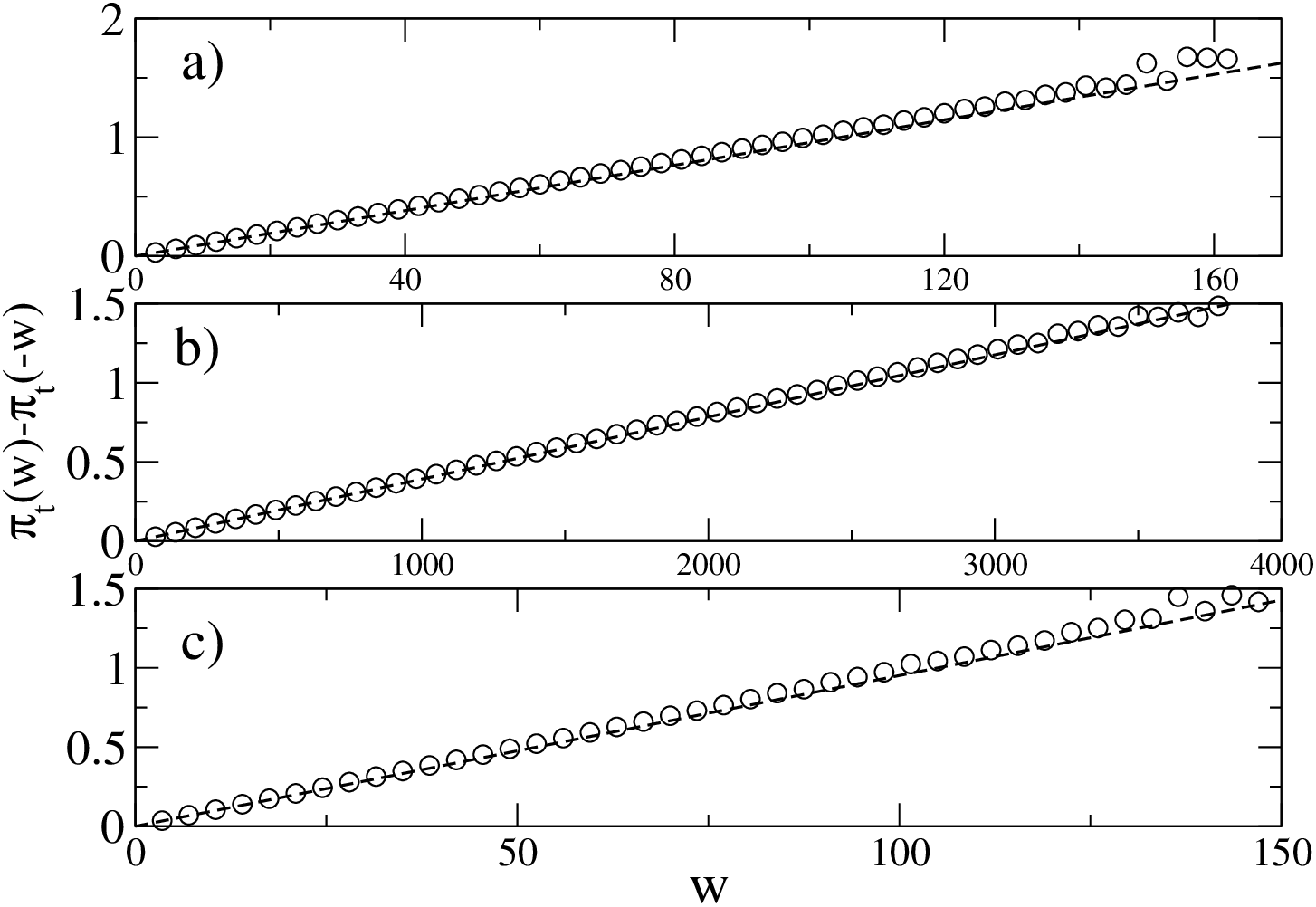}\\
\end{center}
\caption{Test of the Gallavotti-Cohen relation $\pi_t(w)-\pi_t(-w)= 
  \beta_{eff} w$, for the same choices of the parameters as in
  Fig.~\ref{fig:pdf}. The values of the slope $\beta_{eff}$ of the fitting
  dashed lines are in table~\ref{tab:gc}}\label{fig:gc}.
\end{figure}

The comparison with a Gaussian with same mean value and same variance shows
that the pdf $P({\cal W},t)$ is not exactly a Gaussian. In particular there
are deviations from the Gaussian form in the right (positive) tail. This is
well seen in Fig.~\ref{fig:pdfratio}. It must be noted that the largest
deviations in the right tail arise at values of ${\cal W}(t)$ larger than the
minimum ${\cal W}(t)$ available in the left tail, i.e. they have no influence
in the following plot of Fig.~\ref{fig:gc} regarding the Gallavotti-Cohen
symmetry.

\begin{figure}[ht]
\begin{center}
  \includegraphics[height=8cm,clip=true]{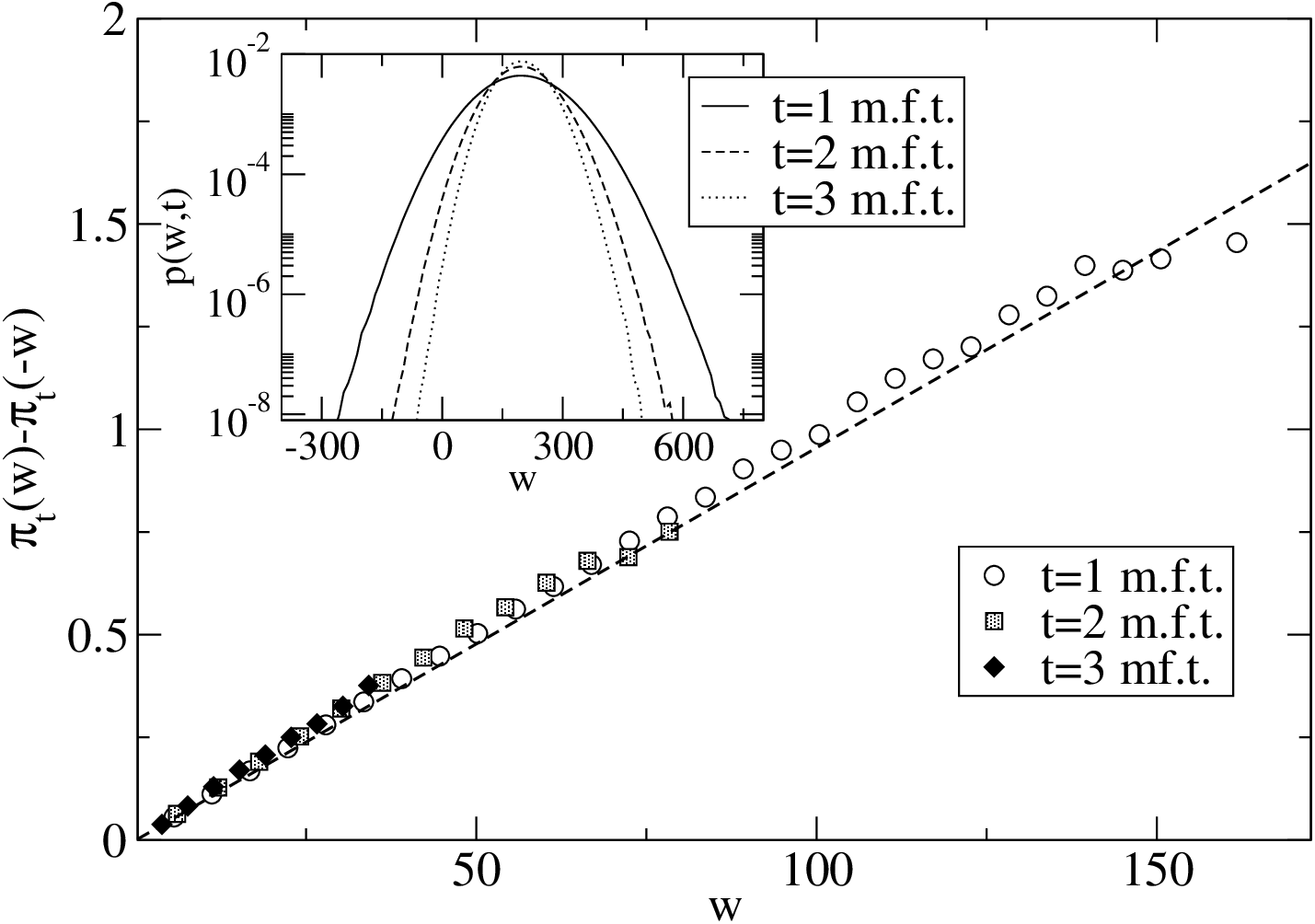}\\
\end{center}
\caption{Test of the Gallavotti-Cohen relation
  $\pi_t(w)-\pi_t(-w)=\beta_{eff} w$, for the system with $N=100$ and
  $\Gamma=0.5$ for different values of $t$. We recall that in this case
  $\langle w \rangle = 100$. The dashed line has slope $\beta=1/T_g$. In the
  inset the corresponding $p(w,t)$ are shown.}\label{fig:gctau}
\end{figure}

In Fig.~\ref{fig:gc}, the finite time Gallavotti-Cohen relation
$\pi_t(w)-\pi_t(-w)=\beta_{eff} w$ is tested for the same choices of the
parameters. The relation, at this level of resolution and for this value of
the time $t$ ($1$ mean free time), is well satisfied.  Moreover
table~\ref{tab:gc} shows that the value of $\beta_{eff}$ is well approximated
by $1/T_g$, as expected if the truncation of $\mu(\lambda)$ at the
second order were valid, see eq.~\eqref{pi_secondorder}. In
Fig.~\ref{fig:gctau} the same relation is checked for different,
slightly larger values of
$t$ (i.e. up to $t$ equal to $3$ mean free times). No
relevant deviations are observed as $t$ is increased.  Moreover this figure is
important to understand the dramatic consequences that larger $t$ have on
the possibility to check numerically 
the Gallavotti-Cohen symmetry: as $t$ is increased,
events with negative integrated power injection become rarer and rarer. This
eventually leads to the vanishing of the left branch of $P({\cal W},t)$.

\begin{table}[ht]
\begin{tabular}{|l|l|l|l|}
  N   &$\Gamma$  &$\beta_{eff}$ &$1/T_g$\\ \hline
  100  &0.5 &0.0100  &0.00955  \\ \hline
  100  &12.5 &0.000402 &0.000382 \\ \hline
  200  &0.5 &0.00995  &0.00952
\end{tabular}
\caption{Factor of proportionality in the ``Gallavotti-Cohen'' relation
  compared with $\beta=1/T_g$.}\label{tab:gc}
\end{table}

\begin{figure}
\begin{minipage}[t]{.46\linewidth}
 \includegraphics[width=1 \textwidth,clip=true]{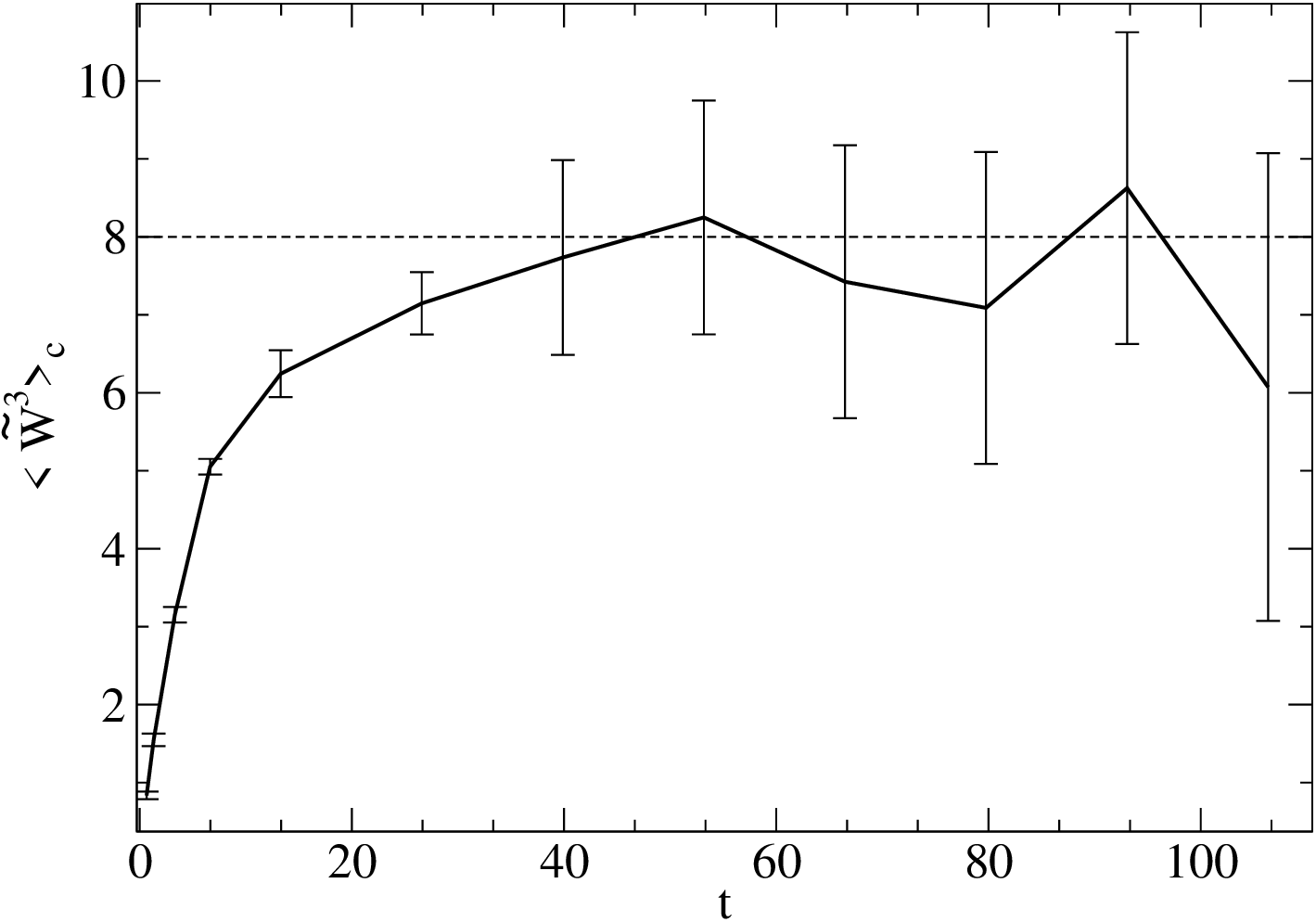}
\caption{\label{thirdcumulant}
  Dimensionless third cumulant $\langle {\tilde\cW}^3 \rangle_c = \langle
  \cW^3\rangle_c/(\langle \cW \rangle T_g^2)$ for several times of
  integration. The time is in units of the mean collision time. Note that the
  time at which a stationary value of the rescaled cumulant is reached is much
  larger than the characteristic time of the system (the collision time).  The
  horizontal dashed line shows the analytical prediction of Eq.
  (\ref{4cumulants}): $\langle \widetilde\cW^3 \rangle =8$.}
\end{minipage}
\hfill
\begin{minipage}[t]{.46\linewidth}
  \includegraphics[width=1 \textwidth,clip=true]{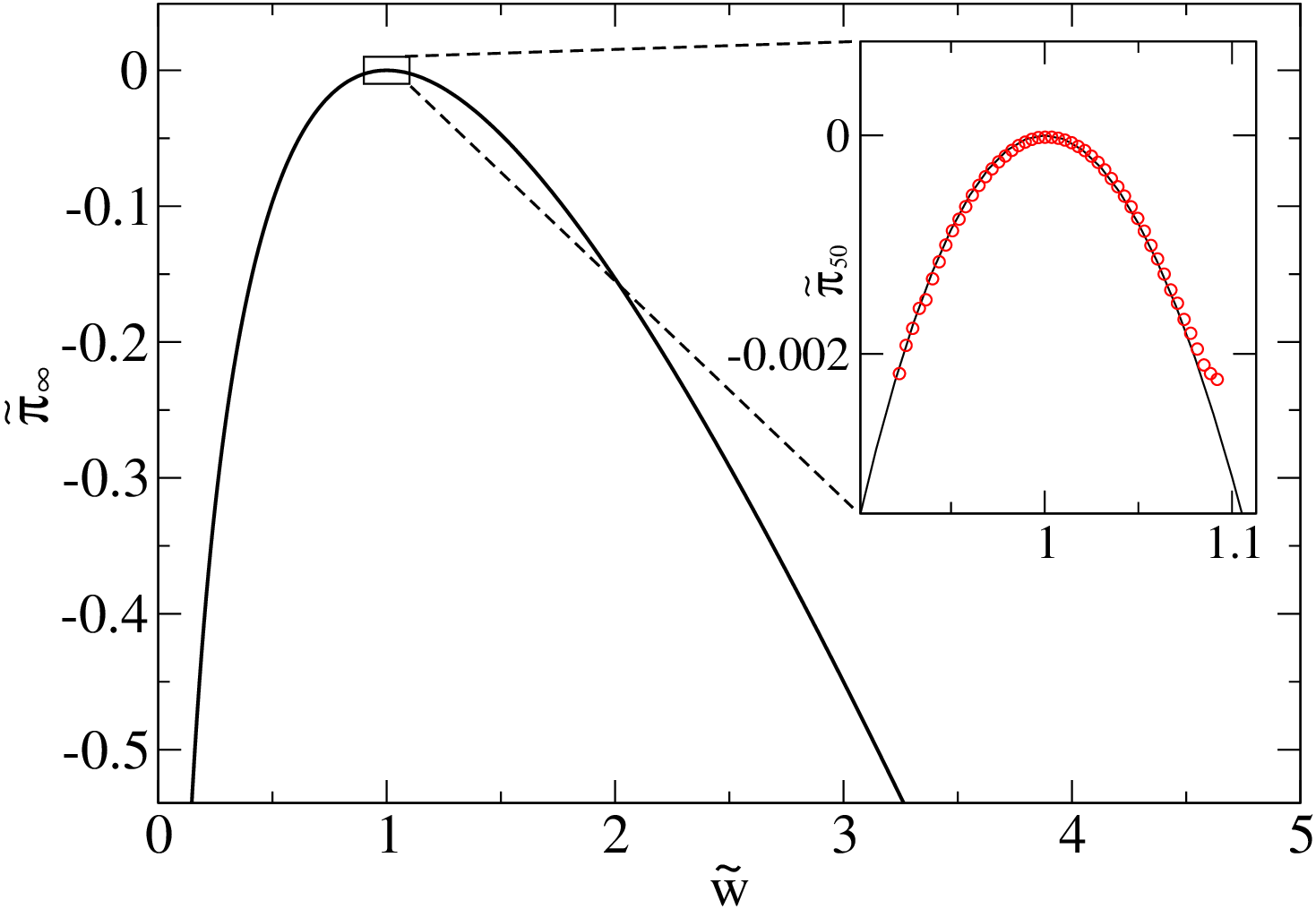}
\caption{\label{piorfwnumeric} Numerical measure of $\pit_t$ for a time of
  50 collisions per particle (when a stationary value for the rescaled
  third cumulant is reached), shown with circles. The full line corresponds to
  $\pi_\infty(w)$ encoded in Eq. (\ref{eq:mudinfini}) and displayed
  in Fig.~\ref{piofw}.}
\end{minipage}
\end{figure}

The main conclusion is that no appreciable departure from the $\lambda^2$
truncation is observed at this level of resolution. Much larger statistics are
required to probe the very high energy tails of $p(w,t)$. Further numerical
insights make evident that the small times used to check the GCFR ($t$
smaller or equal to $3$ mean free times) are far from the time where the
asymptotic large deviation scaling starts working. In
Fig.~\ref{thirdcumulant} indeed, we show the numerically obtained third
cumulant of ${\cal W}(t)$ (rescaled by the first cumulant),
as a function of time. 
The time of saturation is of the order of $\sim 50$ mean
free times. The saturation value is in very good agreement with the value
predicted by our theory, eq.~\eqref{4cumulants}. Note that this value is not
at all trivial, since the third cumulant for a Gaussian distribution is zero.
For this value of $t$, the numerically measured
$\pi_t(w)$ is shown in Fig.~\ref{piorfwnumeric},
rescaled by $\langle w \rangle$. The accessible range of values from a
numerical simulation is dramatically poor and it is already
remarkable to have obtained a good measure of the third cumulant with such a
resolution.

To understand the reason for a verification at small times of the GCFR 
we build upon an argument put forward in~\cite{aumaitrefauvemcnamarapoggi}
where it was noted that for small arguments, $\pi$ can be linearized, 
resulting in a linear behavior of $\pi_t(w)-\pi_t(-w)$. To go further
and compute the prefactor of the linear term, we remark that
near $w=0$ the pdf of $w$ is almost a Gaussian. In the Gaussian
case we immediately get $\pi_t(w)-\pi_t(-w)=\beta_{eff}w$ with $\beta_{eff}=2
\langle {\cal W}(t) \rangle/\langle {\cal W}(t)^2 \rangle_c$. The first two
cumulants at small times are easily obtained considering an uncorrelated
sequence of energy injection, obtaining $\langle {\cal W}(t) \rangle/t
=N\Gamma d$ and $\langle {\cal W}(t)^2 \rangle_c/t=\langle (\sum_i
\mathbf{F}_i^{th} \cdot \mathbf{v}_i)^2 \rangle_c =2N\Gamma d T_g$. 
The value of the slope $\beta_{eff}=1/T_g$ is a direct consequence of these
simple relations. In this case the GCFR observed is
the Green-Kubo-like (or Einstein) relation, which has been investigated
numerically in~\cite{fd} and analytically in~\cite{Garzo},
with the conclusion that such a relation holds within numerically accessible
precision although it is, strictly speaking, invalid \cite{Garzo}.
Small deviations from a Gaussian appear, in
first approximation, as small deviations from the slope $1/T_g$, but the
straight line behavior is robust since the first non-linear term of
$\pi_t(w)-\pi_t(-w)$ is not $w^2$ but $w^3$.

\subsection{Inelastic Maxwell Model}

\begin{figure}[ht]
\begin{center}
\includegraphics[height=8cm,clip=true]{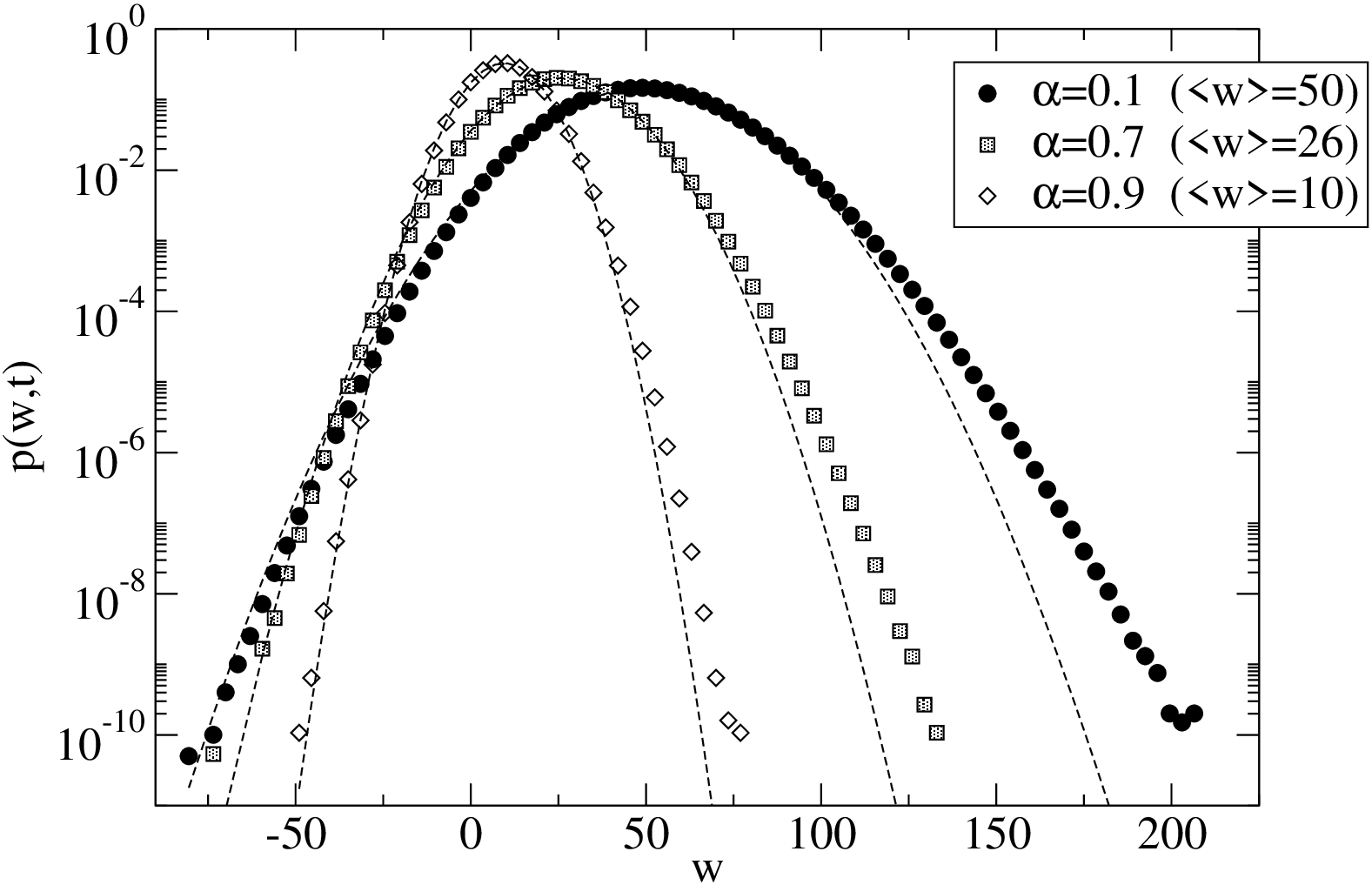}\\
\end{center}
\caption{$p(w,t)\equiv t P(wt,t)$ for different values of $\alpha$ (at fixed
  constant temperature $T_g$) in the Driven Inelastic
  Maxwell Model measured at a time $t$ equal to $1$ mean free time. The dashed
  lines are Gaussian distributions with the same mean and same variance. These
  distributions have been obtained with $\sim 4 \times 10^{10}$ independent
  values of ${\cal W}(t)$.} \label{fig:max:pdf}
\end{figure}

\begin{figure}[ht]           
\begin{center}
 \includegraphics[height=8cm,clip=true]{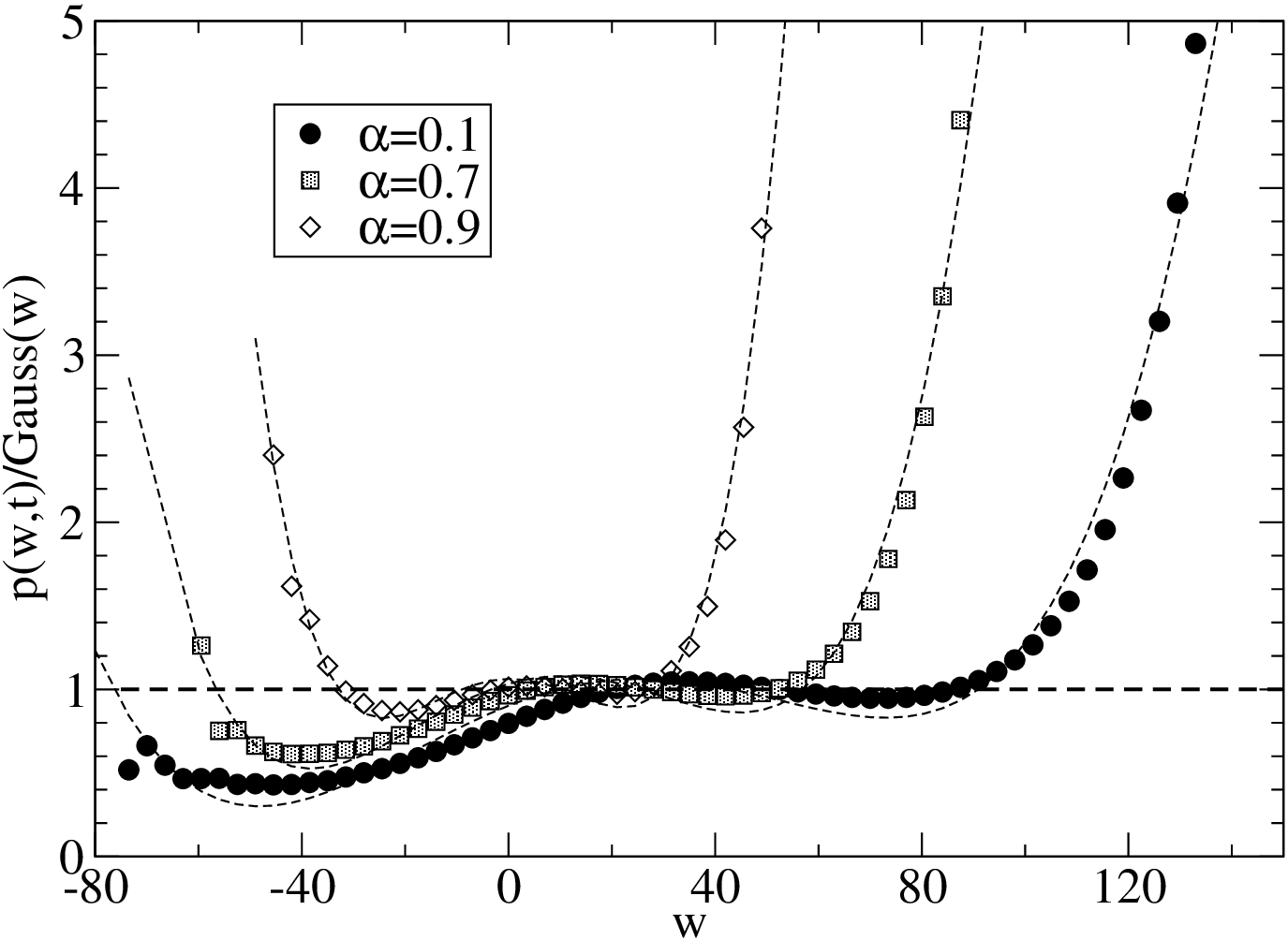}\\
\end{center}
\caption{ \label{fig:max:pdf_ratio}
  $p(w,t)$ (at $t$ equal to $1$ mean free time) divided by a Gaussian
  with same average and same variance for different values of $\alpha$ (at
  fixed constant temperature $T_g$) in the Driven Inelastic Maxwell Model. The
  light dashed lines represent a fit with a polynomial of fourth order.}
\end{figure}

Numerical simulations of the Inelastic Maxwell Model have been performed with a
Direct Simulation Monte Carlo analogous to the one used in the Hard Spheres
model. Thanks to the simplifications present in this model, we are able to
improve the number of collected data by a factor of more than ten.  The
distributions of the injected power $p(w,t)$ are shown in
Fig.~\ref{fig:max:pdf} for some choices of the restitution coefficient
$\alpha$. The driving amplitude $\Gamma$ has been changed in order to keep
constant the stationary granular temperature $T_g$.  In
Fig.~\ref{fig:max:pdf_ratio} we have displayed the deviations from the
Gaussian of $P({\cal W},t)$. The non-Gaussianity of $P({\cal W},t)$ is highly
pronounced, but again it is striking only in the positive branch of the pdf.
We have tried, with success, a fit with a fourth order polynomial.
We observe that the reliability of such a fit increases with $\alpha$,
which is reminiscent of Sonine corrections phenomenology 
for the single particle velocity distribution alluded to earlier.

\begin{figure}
\begin{center}
  \includegraphics[height=8cm,clip=true]{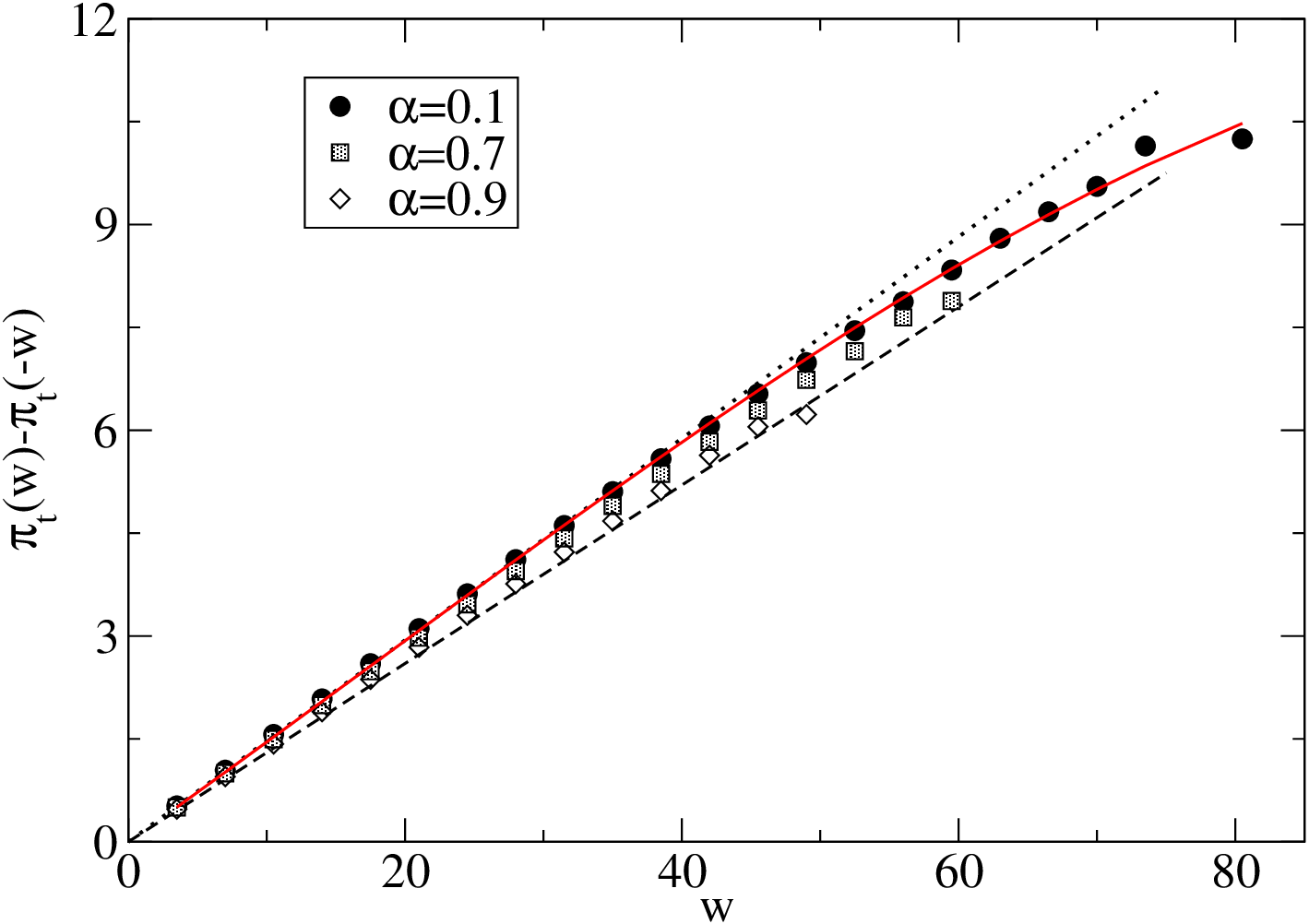}\\
\end{center}
\caption{({\bf Color online}). Finite time test of Gallavotti-Cohen 
  relation for the injected power (with $t$ equal to $1$ mean free time), i.e.
  $\pi_t(w)-\pi_t(-w)$ vs. $w$, in a numerical simulation of the Driven
  Inelastic Maxwell Model with $N=50$, and different values of $\alpha$ (the
  driving amplitude $\Gamma$ has been rescaled in order to fix the granular
  temperature $T_g$).  The dashed curve is a straight line with slope
  $\beta=1/T_g$. The dotted curve is a straight line obtained fitting the
  $\alpha=0.1$ data points until $w=45$, useful as a guide for the eye. The
  thin (red) solid curve is a fit with a cubic ($0.28w + 5.6\cdot 10^{-4} w^2
  - 1.1\cdot 10^{-5} w^3$).} \label{fig:max:gc}
\end{figure}

Finally, in Fig.~\ref{fig:max:gc}, we have attempted a check of the
Gallavotti-Cohen fluctuation relation. The relation seems to be systematically
violated. This appears in two points: 1) the right-left ratio of the large
deviation function is not a straight line; 2) the best fitting line has a
slope which is larger than $\beta$. The ``curvature''(and the deviation from
the slope $\beta$) increases with decreasing values of $\alpha$, indicating
that the inelasticity is the cause of the deviation from the Gallavotti-Cohen
relation. It should be noted that to achieve this result we have collected
more than $4 \times 10^{10}$ independent values of ${\cal W}(t)$, so that the
statistics of the negative large deviations could be clearly displayed.

We also note that this numerically accessible violation of the GCFR is
of a different nature than the one analytically shown in the previous
sections since it occurs at the finite times, i.e. when $\pi_t$ still
displays negative $w$ events.

\section{Conclusion}
\label{sec:conclusion}

In this paper we have given the details of the derivation of an equation for
the cumulants generating function $\mu(\lambda)$ of the injected power in
granular gases, in the case of a homogeneous random driving mechanism. This
equation appears as a generalization of the Boltzmann equation for the
velocity pdf of the gas. It has also the remarkable physical interpretation of
being the kinetic equation of a gas of annihilating/cloning/colliding
particles with an external energy source. This interpretation leads to our
main result: the large deviation function $\pi_{\infty}(w)$ of the injected
power has no negative branch. We have also shown how to exploit a Sonine
expansion to get better and better approximations of $\mu(\lambda)$ near
$\lambda=0$.  We finally obtained $\mu(\lambda)$ in a closed analytical form
by means of a generalized Gaussian hypothesis which is duly justified. This
approximation becomes exact in the large dimension $d \to \infty$ limit.  We
have also presented an argument that suggests the equivalence between the
large deviation function of ${\cal W}(t)$ and that of the energy dissipated in
collisions ${\cal D}(t)$. All these results are consistent with the absence of
negative events in $\pi_{\infty}(w)$. This is at odds with previous studies
that concluded that the GCFR could be satisfied for the integrated injected
power in granular gases.  Numerical simulations, on the other hand, show that
the times at which a check of GCFR can be investigated are much smaller than
the times required to reach the asymptotic large deviation scaling. At such
small times the pdf of ${\cal W}(t)$ near ${\cal W}(t)=0$ is almost Gaussian
and this automatically leads to an artificial verification of the GCFR. On the
other hand, the simulation of the inelastic Maxwell model instead of the
inelastic Hard Spheres model made us able to gather much larger statistics so
that we could see that even at finite times, for which $\pi_{t}(w)$ still
displays negative events, the GC relation can be violated.

We emphasize here that observing non-Gaussian features in the tail of the
distribution of injecting power does not guarantee that the asymptotic scaling
regime has been reached. Non Gaussianities may equally well arise from short
time effects, and it turns out that the latter mechanism is at work already at
very short times. Observing an apparent GCFR at such short times appears to be
an artifact, and much larger times are required to sample correctly the
asymptotic regime.  In this respect, the time behavior of rescaled cumulants
(as displayed in Fig. \ref{thirdcumulant}) is a valuable tool to decide if the
``transient'' regime is over.

The analytical method put forward here to compute large temporal
deviations is quite general and is particularly relevant in the
context of the computation of probability distributions of global
observables, which are useful, as underlined in the introduction,
for the determination of universal features of non-equilibrium
systems. For example, it applies to ballistically controlled processes,
be they at equilibrium or not. Future work along these lines
also includes the consideration of out of equilibrium granular 
gases without a steady state, such as freely cooling systems.

\begin{acknowledgments} A. P. acknowledges the Marie Curie grant No.
MEIF-CT-2003-500944. E.T. thanks the EC Human Potential program under
contract HPRN-CT-2002-00307 (DYGLAGEMEN).
\end{acknowledgments}

\addcontentsline{toc}{section}{References}
\bibliographystyle{hunsrt}
\bibliography{notes}

\newpage
\begin{thebibliography}{0}
  
\bibitem{vankampen} N. van Kampen, {\it Stochastic processes in physics and
  chemistry}, North-Holland, 1992.

\bibitem{einstein} A. Einstein, Ann. d. Phys. {\bf 17}, 549 (1905).
  
\bibitem{onsager} L. Onsager, Phys. Rev. {\bf 37}, 405 (1931) ; Phys. Rev. {\bf 38},
  2265 (1931).
  
\bibitem{green} M.S. Green, J. Chem. Phys. {\bf 22}, 398 (1954).
  
\bibitem{kubo} R. Kubo, J. Phys. Soc. Japan {\bf 12}, 570 (1957).

\bibitem{degroot} S. R. de Groot and P. Mazur, {\it Non-equilibrium
    thermodynamics}, North-Holland, 1969.

\bibitem{pinton} S.T. Bramwell, P.C.W. Holdsworth and J.-F. Pinton,
Nature {\bf 396}, 552-554, (1998).
  
\bibitem{brey} J. Javier Brey, M. I. Garc\'{\i}a de Soria, P. Maynar, and M. J.
Ruiz-Montero, Phys. Rev. Lett. {\bf 94}, 098001 (2005)







\bibitem{evanscohenmorriss} D.J. Evans, E.G.D. Cohen and G.P. Morriss, Phys.
  Rev. Lett. {\bf 71}, 2401 (1993).

\bibitem{gallavotticohen} G. Gallavotti and E.G.D. Cohen, Phys. Rev. Lett.
  {\bf 74}, 2694 (1995).


\bibitem{thermostats} D. J. Evans and G.P. Morriss, {\it Statistical Mechanics
    of Nonequilibrium Liquids}, Academic Press, London, 1990; G. P. Morriss
  and C. P. Dettmann, Chaos {\bf 8}, 321 (1998).

\bibitem{kurchan} J. Kurchan, { J. Phys. A} {\bf 31}, 3719 (1998).

\bibitem{lebowitzspohn} J.L. Lebowitz and H. Spohn, J. Stat. Phys. {\bf 95},
  333 (1999).
  
\bibitem{experiments} S. Ciliberto, and C. Laroche, Journal de Physique IV,
  {\bf 8} , 215 (1998); N. Garnier and S. Ciliberto, Phys. Rev. E {\bf 71}
  060101(R) (2005); S. Ciliberto, N. Garnier, S. Hernandez, C. Lacpatia, J.-F.
  Pinton and G.  Ruiz Chavarria, Physica A {\bf 340} (1-3) pp 240-250 (2004).

\bibitem{aumaitrefauvemcnamarapoggi} S. Auma\^\i tre, S. Fauve, S. McNamara and
  P.  Poggi, Eur. Phys. J. B {\bf 19}, 449 (2001).

\bibitem{farago} J. Farago, J. Stat. Phys. {\bf 107}, 781 (2002).

\bibitem{feitosamenon} K. Feitosa and N. Menon, Phys. Rev. Lett. {\bf 92},
  164301 (2004).

\bibitem{letter2} P. Visco, A. Puglisi, A. Barrat, E. Trizac and F. van
  Wijland, accepted in Europhysics Letters (2005).
  
\bibitem{letter1} A. Puglisi, P. Visco, A. Barrat, E. Trizac and F. van
  Wijland, submitted (2005).

\bibitem{kicks} D. R. M. Williams and F. C. MacKintosh, Phys.
  Rev. E {\bf 54} R9 (1996);  G. Peng and T. Ohta,
  Phys. Rev. E {\bf 58}, 4737 (1998); T.P.C. van Noije, M.H. Ernst, E. Trizac
  and I. Pagonabarraga, Phys. Rev E {\bf 59}, 4326 (1999); C. Henrique, G.
  Batrouni and D. Bideau, Phys. Rev. E {\bf 63}, 011304 (2000); S. J. Moon, M.
  D. Shattuck and J. B. Swift, Phys. Rev. E {\bf 64}, 031303 (2001); I.
  Pagonabarraga, E. Trizac, T.P.C. van Noije and M.H. Ernst, Phys. Rev. E {\bf
    65}, 011303 (2002).

\bibitem{prevost}
A. Prevost, D.A. Egolf, and J.S. Urbach,
Phys. Rev. Lett. {\bf 89}, 084301 (2002).

\bibitem{vannoijeernst} T.P.C. van Noije and M.H. Ernst, Granular Matter {\bf
    1}, 57 (1998).

\bibitem{montanerosantos} J.M. Montanero and A. Santos, Granular Matter {\bf
    2}, 53 (2000).

\bibitem{coppex} F. Coppex,   M. Droz,   J. Piasecki and  E. Trizac,  
Physica A {\bf 329}, 114 (2003).

\bibitem{bird} G. A. Bird, Molecular Gas Dynamics and the Direct
Simulation of Gas Flows, Clarendon 1994 (Oxford).

\bibitem{farago2} 
J. Farago, Physica A {\bf 331}, 69 (2004).
 
\bibitem{vanzoncohen} 
R. van Zon and E. G. D. Cohen, 
Phys. Rev. Lett. {\bf 91}, 110601 (2003).

\bibitem{feller} W. Feller, {\it An Introduction to Probability Theory and Its
    Applications}, John Wiley \& Sons 1966.

\bibitem{maxwell} J. C. Maxwell, Phil. Trans. {\bf 157}, 49 (1867).
  
\bibitem{puglisi} A. Baldassarri, U. Marini Bettolo Marconi and A. Puglisi,
  Europhys. Lett. {\bf 58}, 14-20 (2002).

\bibitem{bennaimkrap} E. Ben-Naim and P.L. Krapivsky, J. Phys. A {\bf
    35}, L147 (2002); Lecture Notes in Physics {\bf 624}, 65 (2003).
  
\bibitem{britoernst} M. H. Ernst and R. Brito, Europhys. Lett. {\bf 58}, 182
  (2002).
  
\bibitem{bobylev} A. V. Bobylev, J. A. Carrillo, I. M. Gamba, J. Stat. Phys.
 {\bf 98}, 743 (2000).
  
\bibitem{santos} A. Santos, 
Physica A {\bf 321}, 442 (2003).

\bibitem{fd} A. Puglisi, A. Baldassarri and V. Loreto, 
Phys. Rev. E {\bf 66}, 061305 (2002).

\bibitem{Garzo}
V. Garz\'o, Physica A {\bf 343}, 105 (2004).
  
\end{thebibliography}

\end{document}